\documentclass[10pt,twocolumn]{IEEEtran}
\usepackage{amsmath}
\usepackage{amsfonts}
\usepackage{amssymb}
\usepackage{dsfont}
\usepackage{enumitem}
\usepackage{multirow}
\usepackage{color}
\usepackage{graphicx}
\usepackage{epstopdf}
\usepackage{caption}
\usepackage{cases}
\usepackage{subfigure}
\usepackage{cite}
\usepackage{bm}
\usepackage{algorithm}
\usepackage{algorithmicx}
\usepackage{algpseudocode}
\usepackage{setspace}

\newtheorem{lemma}{Lemma}
\newtheorem{remark}{Remark}
\newtheorem{theorem}{Theorem}
\newtheorem{corollary}{Corollary}
\newtheorem{proposition}{Proposition}
\begin{document}

\title{LPD Communication: A Sequential Change-Point Detection Perspective}
\author{ Ke-Wen Huang, Hui-Ming Wang, \emph{Senior Member, IEEE}, \\
Don Towsley, \emph{Life Fellow, IEEE},
and H. Vincent Poor, \emph{Fellow, IEEE}
\thanks{\scriptsize
K.-W. Huang and H.-M. Wang are with the School of Information and Communication Engineering, Xi'an Jiaotong University, Xi'an 710049, Shaanxi, P. R. China
(e-mail: {\tt
xjtu-huangkw@outlook.com, xjbswhm@gmail.com}).
}
\thanks{\scriptsize
D. Towsley is with the College of Information \& Computer Sciences, University of Massachusetts, Amherst, MA 01003 USA (e-mail: {\tt towsley@cs.umass.edu}).}
\thanks{\scriptsize
H. V. Poor is with the Department of Electrical Engineering, Princeton University, Princeton, NJ 08540 USA (email: {\tt poor@princeton.edu}).}
}

\IEEEtitleabstractindextext{
\begin{abstract}
In this paper, we establish a framework for low probability of detection (LPD) communication from a \emph{sequential change-point detection} (SCPD) perspective, where a transmitter, Alice, wants to hide her transmission to a receiver, Bob, from  an adversary, Willie. The new framework facilitates modeling LPD communication and further evaluating its performance  under the condition that Willie has no prior knowledge about when the transmission from Alice might start and that Willie wants to determine the existence of the communication as quickly as possible in a real-time manner.
We consider three different sequential tests, i.e., the Shewhart, the cumulative sum (CUSUM), and the Shiryaev-Roberts (SR) tests, to model Willie's detection process.
Communication is said to be covert if it ceases before being detected by Willie with high probability.
Covert probability defined as the probability that Willie is not alerted during Alice's transmission is investigated.
We formulate an optimization problem aiming at finding the transmit power and transmission duration so as to maximize the total amount of information that can be transmitted subject to a high covert probability.
Under the Shewhart test, closed-form approximations of the optimal solutions are derived, which well approximate the solutions obtained from exhaustive search.
As for the CUSUM and SR tests, we provide effective algorithms to search for the optimal solutions.
Numeric results are presented to show the performance of LPD communication.
\end{abstract}
}
\maketitle

\IEEEdisplaynontitleabstractindextext
\IEEEpeerreviewmaketitle

\section{Introduction}
Recently, \emph{low probability of detection (LPD)} communication, also referred to as \emph{covert} communication, has gained considerable attention \cite{A.O.Hero,B.A.BashCM2015,S.Yan219Arxiv}.
The premise behind LPD communication is to make the existence of the communication undetectable to third parties.
Consider a scenario where a transmitter, Alice, transmits to a receiver, Bob, under the surveillance of an adversary, Willie, who aims at determining whether communication is ongoing between Alice and Bob.
Using the technique of LPD communication, even though Willie keeps monitoring the channel, it cannot make an effective decision on whether Alice is on transmission, which is able to to protect the privacy of Alice and Bob \cite{S.Yan219Arxiv}.

\subsection{Related works}
The definition and fundamental information-theoretical performance limits of LPD communication was firstly proposed in \cite{B.A.BashJSAC2013}, and further investigated in \cite{P.H.Che2013,B.A.BashJSAC2013,L.WangTIT2016,M.R.BlochTIC2016,S.YanTWC2019Gau}.
In \cite{B.A.BashJSAC2013}, the sum of false alarm and miss detection probabilities was defined as the error probability of Willie's detector, and LPD communication performance limits subject to an arbitrarily high error probability approaching one was investigated. It was proved in
\cite{B.A.BashJSAC2013} that
LPD communication in additive white Gaussian noise (AWGN) channel obeys the \emph{square root law} (SRL), i.e., Alice can transmit at most $O(\sqrt{n})$ bits covertly and reliably to Bob in $n$ channel usages. Attempting to break the SRL leads to the consequence that either the communication can be detected by Willie, or Bob is not able to perform error-free decoding.
In \cite{P.H.Che2013}, it was shown that the SRL also holds in a binary symmetric channel (BSC) with no common randomness if the channel between Alice and Bob is better than that between Alice and Willie.
In \cite{L.WangTIT2016}, the authors extended the SRL to general discrete memoryless channels (DMCs), and characterized the expressions of the scaling constants in DMC and AWGN channel. In \cite{M.R.BlochTIC2016}, it was further  revealed that the required length of the shared secret key to enable LPD communication was $O(\sqrt{n})$ bits for general DMCs.
The authors of \cite{S.YanTWC2019Gau} investigated the optimal signalling for LPD communication in AWGN channel. It was revealed in \cite{S.YanTWC2019Gau} that if $\mathbb{D}(p_0||p_1)$ is used as a metric of covertness, then Gaussian signalling is optimal, however, this is not the case if $\mathbb{D}(p_1||p_0)$ is used as a metric of covertness, where $p_1$ and $p_0$ denote distributions of the signal received by the adversary when LPD communication exists and does not exist, respectively, and $\mathbb{D}(\cdot||\cdot)$ represents the Kullback-Leibler divergence.

The SRL in LPD communication is discouraging as it implies that the covert information rate tends to be zero as the number of channel usages increases. Many recent works  \cite{B.A.BashISIT2014,B.A.BashTWC2016,K.S.K.Arumugam2016,D.Goeckel2016CL,T.V.SobersTWC2017} tried to improve it and revealed some scenarios where the SRL could be broken and a positive covert information rate might exist, for examples, when Willie does not exactly know when Alice starts to transmit (if she does)  \cite{B.A.BashISIT2014,B.A.BashTWC2016,K.S.K.Arumugam2016},
when Willie is uncertain about the noise variance on its channel \cite{D.Goeckel2016CL},
and when a third party broadcasts jamming signals \cite{T.V.SobersTWC2017}.
Under the above fundamental information-theoretical framework of LPD communication, some practical scenarios have been investigated,
e.g., in point-to-point communication systems \cite{S.LeeJSTSP2015,B.HeCL2017,S.YanICC2017,S.YanITFS2019,F.Shu2019WCL},  in relay systems \cite{A.Sheikholeslami2018TWC,J.Hu2019TWC}, and in random networks \cite{B.HeTWCtobepublished,T.-X.ZhengTWC2019}.
Specifically, in \cite{S.LeeJSTSP2015} and \cite{B.HeCL2017}, the covert transmission rate maximization problem was investigated, where Willie was assumed to have inaccurate knowledge about the noise power.
By taking the effect of finite block length into consideration, \cite{S.YanICC2017} and \cite{S.YanITFS2019} maximized the effective covert throughput subject to a lower bound on the detection error probability of Willie.
The scenario investigated in \cite{S.YanITFS2019} was further extended to the case with a full-duplex receiver in \cite{F.Shu2019WCL}, and LPD communication performance was improved by letting the receiver broadcast artificial noise.
In \cite{J.Hu2019TWC}, LPD communication  was considered in an one-way relay network, wherein the relay node harvested energy from the source node and tried to transmit its own message covertly.
In \cite{A.Sheikholeslami2018TWC}, the authors proposed to use multiple relays to forward the private message covertly, and provided algorithms to find the route that maximizes the covert throughput or minimizes end-to-end delay.
In \cite{B.HeTWCtobepublished,T.-X.ZhengTWC2019}, the authors studied the performance of LPD communication in the presence of randomly located interferers.

\subsection{Binary hypothesis test framework in existing works}
Though existing works have provided us considerable understanding on achievable performance of LPD communication, they have two major limitations.

\begin{figure*}[t]
  \centering
  \includegraphics[width=5.5 in]{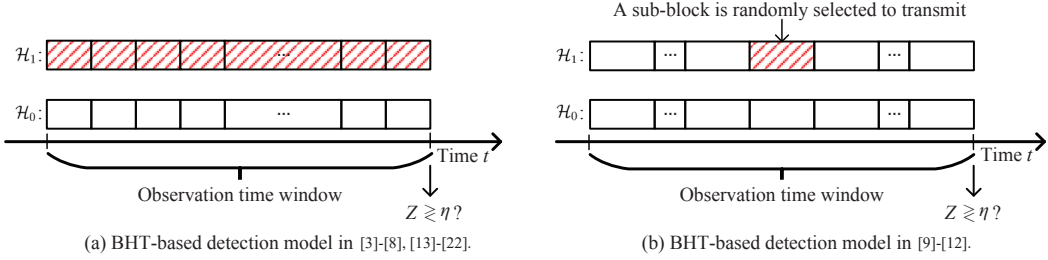}
  \caption{\small The BHT-based detection model in existing works, where $Z$ is the detection statistic.}\label{BHTBasedDetection}
  \vspace{-3mm}
\end{figure*}

First, it was implicitly assumed that the transmission from Alice to Bob either is everlasting all the time or
does not exist at all during Willie's whole detection process. In other words,
Willie's detection was modeled as a binary hypothesis test (BHT) problem, and the following two hypothesises were taken into consideration,
\begin{enumerate}
\item  $\mathcal{H}_0$: communication does not exist, i.e., all observations of Willie are pure thermal noise;
\item  $\mathcal{H}_1$: communication exists, i.e., all observations of Willie are the superposition of Alice's signal and thermal noise.
\end{enumerate}
All the works in \cite{P.H.Che2013,B.A.BashJSAC2013,L.WangTIT2016,M.R.BlochTIC2016,T.V.SobersTWC2017,S.YanTWC2019Gau,B.HeTWCtobepublished,
	A.Sheikholeslami2018TWC,S.YanITFS2019,S.YanICC2017,S.LeeJSTSP2015,B.HeCL2017,F.Shu2019WCL,J.Hu2019TWC} discussed covert communication under such a detection framework, which is illustrated in Fig . \ref{BHTBasedDetection}(a).
However, in practice, it is likely that the Alice's transmission starts at some time in the  \emph{middle} of Willie's observation procedure. This is because Alice and Bob determine when their communication starts, which is unknown to Willie. In this case, such a BHT-based model cannot describe the behavior of LPD communication accurately due to the \emph{asynchrony} between Alice and Willie.

Another issue is that BHT-based detection works in batch mode, namely, all sample observations are collected before making a decision.
But in practice, Willie usually does not have to wait until all data have been collected to make the decision off-line.
Instead, he can make the decision while successively observing.
In fact, as a surveillant, Willie should perform the detection in a \emph{sequential} manner, which is a very natural requirement since in many applications decisions should be made in real time.

We note that a slightly different model, as illustrated in Fig. \ref{BHTBasedDetection}(b),  was investigated in \cite{B.A.BashISIT2014,B.A.BashTWC2016,K.S.K.Arumugam2016}.
Specifically, in \cite{B.A.BashISIT2014,B.A.BashTWC2016,K.S.K.Arumugam2016},
the communication between Alice and Bob occurs during a finite block of time, the beginning of which is random and unknown to Willie.
This model indeed has taken the asynchrony between Alice and Willie into consideration.
However, for Willie's detection, they still assume that all the observations were collected before making the final decision  by performing a BHT.

\subsection{Motivations and contributions}
Based on the above observations,
in this paper, we establish a new framework for LPD communication under the following two basic and important assumptions:
\begin{enumerate}
\item Willie has no prior knowledge about when the communication between Alice and Bob may start;
\item Willie performs a sequential detection to discover the communication.
\end{enumerate}

To be more specific,  we view the received signal sequence of Willie as a discrete time stochastic process.
If there is no communication between Alice and Bob, then Willie's received signal only consists of thermal noise.
If there exists a time point (unknown to Willie) at which Alice starts to transmit to Bob, then after this moment,  Willie's received signal becomes the superposition of the information symbol transmitted by Alice and the thermal noise.
In other words, there exists a change of statistic feature of Willie's received signal sequence.
From the perspective of Willie, determining whether communication occurs between Alice and Bob is equivalent to determining whether such a change occurs.
As Willie has no idea when the communication occurs, the detector should work in a sequential manner, i.e., at each time a new sample observation is obtained, it decides whether or not communication occurs based on all the samples obtained so far.

In view of this procedure, the design of Willie's detector naturally falls into the field of \emph{sequential change-point detection} (SCPD) \cite{A.G.Tartakovsky2013}, which is also referred to as \emph{quickest detection} \cite{H.V.Poor2009}.
In SCPD, at each time when a new sample observation is obtained, the detector makes a decision based on all the sample observations collected at hand.
If it indicates that no change has occurred, then the detector moves to the next time point at which a new sample observation is obtained and a new decision will be made.
The detection procedure does not stop until the detection result indicates that a change has occurred. We can see that SCPD   naturally works in an on-line manner.


Under the SCPD framework, the performance metric is significantly different  from BHT-based detection.
In BHT-based detection, the performance is fully characterized by \emph{false alarm} and \emph{miss detection} probabilities.
However, in SCPD, \emph{frequency of false alarms} (or the average time interval between two consecutive false alarms) and \emph{average detection delay}, i.e., the average time interval between the time when the change is detected and the time when the change truly occurs, are the primary metrics.
This is because in BHT, only a single decision is made, and therefore, we only need to check whether or not the decision is correct.
However, in SCPD, at each time when a new sample observation is obtained, a new decision will be made, the impact of which is two-fold:
\begin{enumerate}
\item  even if no change occurs, the detector will always raise a false alarm after observing for a sufficiently long time due to the random noise. In fact, many practical SCPD algorithms raise a false alarm with probability one within finite time \cite{H.V.Poor2009,A.G.Tartakovsky2013};
\item if a change occurs at an unknown time point, to guarantee the reliability of the detection result, there usually exists an inevitable detection delay, i.e., there is a gap between when the change occurs and when the change is successfully detected.
\end{enumerate}
Intuitively, if the change is significant,  it is reasonable to expect a small detection delay. However, if the change is small, it will take a long time to successfully detect it.
In reality, we often desire a low rate of false alarm and a short detection delay which, however, cannot be attained at the same time. Therefore, practical SCPD usually achieves a trade-off between the two performance metrics.
Typical SCPD algorithms, such as the Shewhart test, the cumulative sum (CUSUM) test, and the Shiryaev-Roberts (SR) test,
are designed to minimize the detection delay subject to a sufficiently long interval between two consecutive false alarms \cite{A.G.Tartakovsky2013,H.V.Poor2009}.

Based on this critical feature, when Willie performs a sequential test, \emph{it is possible for Alice and Bob to achieve LPD communication by taking advantage of Willie's detection delay}. More specifically, if Alice transmits at a low power level, then it may take a long time for Willie to make a reliable decision. Consequently, if the time duration of communication is less than the detection delay, then Willie will not be alerted with a high probability. In this case, we say that LPD communication succeeds.

From the perspective of Alice and Bob, they want to communicate as much information as possible without being detected by Willie.
Naturally, the above discussion raises a fundamental question:
how much power could Alice use
and how long the time could she exploit to achieve LPD communication?
These questions are the critical concerns of effective LPD communications, which constitute the main topic of this paper.
The novelties and contributions of this paper are summarized below.
\begin{enumerate}
\item We establish a new framework for LPD communication based on SCPD. In our framework, Willie adopts SCPD to detect the occurrence of communication between Alice and Bob in a real-time manner, and Alice and Bob achieve LPD communication by taking advantage of the detection delay of Willie's detector.
    To the best of our knowledge, this is the first paper to investigate the performance of LPD communication under SCPD.

\item Several well-known SCPD algorithms are adopted to model Willie's detection, i.e., the Shewhart test, the CUSUM test, and the SR test. They  are all optimal SCPD detectors under different conditions. For each test, we analyze the covert probability, defined as the probability that Willie is not alerted during the communication period, which depends on both the transmit power and transmission duration.

\item We formulate the fundamental LPD communication problem of maximizing the amount of information transmitted from Alice to Bob subject to a sufficiently high covert probability, wherein the optimization variables are transmit power and transmission duration. For the Shewhart test, we derive a closed-form approximate solution, which achieves similar performance compared to the optimal solution obtained from an exhaustive numeric search. For CUSUM and SR tests,we provide methods to search the optimal solutions.
    Numerical results are presented to show LPD communication performance in our new framework.
\end{enumerate}

\begin{figure*}[t]
  \centering
  \includegraphics[width=5.5 in]{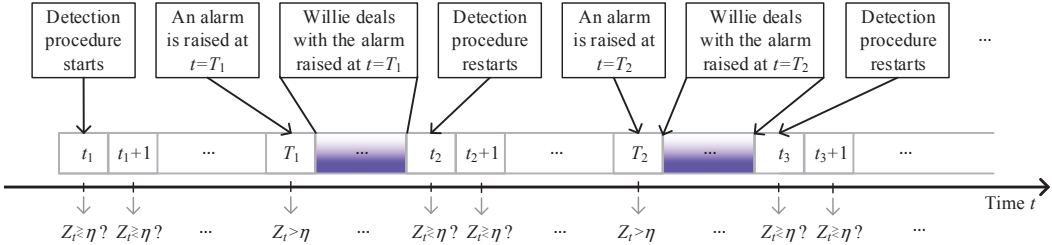}
  \caption{\small The SCPD process of Willie considered in this paper, where the blue shaded regions denote that Willie is dealing with the alarms raised by his detector. Note that for illustrative simplicity, at each time when an alarm is raised, we let Willie fist handle the alarm and then restart the detection. However, in practice, Willie may also directly restart the detection process once an alarm is raised.}\label{SCPDModel}
  \vspace{-3mm}
\end{figure*}

\subsection{Organization and notations}
The rest of the paper is organized as follows:
Section II introduces the system model in detail, including Willie's sequential detection process;
Section III presents covert probabilities and their analysis;
Section IV presents the method to solve the covert throughput maximization problem.
Section V numerically evaluates LPD communication performance;
and finally, Section VI concludes the paper.

\emph{Notation:}
$\mathbb{P}( \cdot )$ and $\mathbb{E}( \cdot )$ denote probability and mathematical expectation, respectively.
$\mathcal{CN}(\mu,\sigma^2)$ represents the complex Gaussian distribution with mean $\mu$ and variance $\sigma^2$.
$\mathcal{E}(\lambda)$ denotes the exponential distribution with parameter $\lambda$.
$a\wedge b$ and $a\vee b$ denote $\min\{a,b\}$ and $\max\{a,b\}$, respectively.
The supremum and infimum  are denoted by $\mathrm{sup}$ and $\mathrm{inf}$, respectively. $\mathbb{I}\{\cdot\}$ is the indicator function.

\section{System model}
We consider LPD communication from Alice to Bob, i.e., Alice transmits a message to Bob while keeping Willie unaware of the transmission.
Let $\nu$ ($\nu = 0,1,2,\cdots$) denote the time  Alice starts to transmit.
Then, the received signal of Willie at time $t$ is
\begin{align}
y_t \triangleq \mathbb{I}\left\{ \nu < t \leq \nu + L \right\}
\sqrt{q}s_t + z_t , \quad t = 1,2,\cdots, \label{ReceivedSignalSequenceModified}
\end{align}
where $z_t\sim\mathcal{CN}\left(0,\sigma_W^2\right)$ denotes the additive white Gaussian noise,
$s_t\sim\mathcal{CN}\left(0,1\right)$ denotes the information symbol transmitted by Alice with
$q$ the transmit power, and
$L\geq 1$ is the transmission duration.
Here, if $\nu = \infty$, it means that Alice does not transmit anything and there is no transmission.
In the case $\nu = 0$, transmission starts at the same time that Willie starts his detection.
According to \eqref{ReceivedSignalSequenceModified}, the probability density function (PDF) of $y_t$  for $t \leq \nu$ or $t > \nu + L$  is $f_{0}(x) \triangleq  \frac{1}{\pi\sigma_W^2} \exp\left(-\frac{|x|^2}{\sigma_W^2}\right)$, and that of $y_t$ for $\nu < t \leq  \nu + L$ is $f_1(x) \triangleq \frac{1}{\pi(q+\sigma_W^2)} \exp\left(-\frac{|x|^2}{q+\sigma_W^2}\right)$.
In the following, for notational simplicity, we assume that $\sigma_W^2 = 1$
\footnote{ For the case $\sigma_W^2 \neq 1$, we can equivalently view $\{\hat{y}_t : t\geq 1 \}$ as the received signal sequence, where $\hat{y}_t = y_t/\sigma_W$. In this way, the equivalent noise $\hat{z}_t$ satisfies that $\hat{z}_t = z_t/\sigma_W\sim\mathcal{CN}(0,1)$. Note that in our case, $\hat{y}_t$ is a sufficient statistic of $y_t$, and therefore this causes no harm to the detection performance of Willie.}.

The proposed signal model in \eqref{ReceivedSignalSequenceModified} can be viewed as a generalization of the model investigated in
\cite{
B.A.BashJSAC2013,P.H.Che2013,S.YanTWC2019Gau,
L.WangTIT2016,M.R.BlochTIC2016,T.V.SobersTWC2017,B.HeTWCtobepublished,A.Sheikholeslami2018TWC,S.YanITFS2019,S.YanICC2017,
S.LeeJSTSP2015,B.HeCL2017,F.Shu2019WCL,J.Hu2019TWC}. Specifically, in these works, if communication between Alice and Bob exists, $\{y_1,y_2,\cdots\}$ are i.i.d. random variables with PDF $f_1(x)$, which can be viewed as a special case of \eqref{ReceivedSignalSequenceModified} with $\nu = 0$. Note that in our model \eqref{ReceivedSignalSequenceModified},  $\nu$ can be an arbitrary non-negative integer.
In our case, Willie's detection procedure cannot be modeled as a BHT problem as in existing works.
This is because $\nu$ being unknown precludes Willie from pre-determining how many samples are needed to perform the BHT as in Fig. \ref{BHTBasedDetection}.
In practice, Willie may want to discover the transmission from Alice as soon as possible once it starts.
This requires a real-time detection procedure for Willie where he continuously checks if Alice is transmitting.
Based on this consideration, in this paper,  Willie adopts an on-line detection procedure that starts at time $t=1$ without loss of generality. Denote by $Z_t$ Willie's test statistic at time $t$, which depends on the sample observations up to time $t$, i.e., $\{y_1,y_2,\cdots,y_t\}$.
Based on $Z_t$, Willie determines whether Alice is transmitting. We next introduce Willie's detection procedure.


\subsection{Adversary model}
\label{IntroductionST}
In this paper, we assume Willie has no prior knowledge about the change-point $\nu$, and therefore, we consider non-Bayesian SCPD at Willie \cite{H.V.Poor2009}
. Theoretically,  SCPD can be described as a \emph{stopping time} $T$ based on the sequential observations $\{y_1,y_2,\cdots,\}$ at which a decision on the occurrence of a change is made. At time $T$, an alarm will be raised. The stopping time $T$ can be written in the form of
\begin{align}
T =\inf\left\{ t : Z_t \geq \eta , t \geq 1 \right\},\label{GeneralizedStoppingTimeForm}
\end{align}
where $Z_t$ is the test statistic at time $t$, which is usually a function of $\{y_1,y_2,\cdots,y_t\}$, and $\eta$ is a pre-designed detection threshold.
According to \eqref{GeneralizedStoppingTimeForm}, $T$ is the first time when the test statistic $Z_t$ exceeds the detection threshold $\eta$. An SCPD process is a sequential test that successively compares $Z_t$ with $\eta$ until $Z_t$ exceeds $\eta$.
SCPD produces one of the following two results:
1) $T>\nu$, which means that the change-point is successfully detected with  the \emph{detection delay} as $T_d \triangleq T - \nu$, and 2) $T \leq  \nu$, which means that a false alarm occurs.
In this paper, we assume that the detection procedure works in a cyclic manner, i.e., at each time when an alarm is raised, which is either a false alarm or a successful detection, Willie   takes some action to deal with it,
and then restarts the detection procedure. The whole detection process of Willie is comprehensively illustrated in Fig. \ref{SCPDModel}.
Note that how Willie responds to the alarms depends on the specific application scenario, and an example will be presented later.

In practice, while avoiding frequently raising false alarms, the detection process should be designed to have a small detection delay
so that timely action can be taken to deal with the occurrence of a change.
The false alarm frequency is usually characterized by the \emph{average run length to false alarm} (ARL2FA), which is defined as $t_a\left(T\right) \triangleq \mathbb{E}\left(T| \nu = \infty\right)$, and the design of the stopping time $T$ (including the construction of $Z_t$ and the value of $\eta$) is subject to a lower bound on the ARL2FA, i.e.,
\begin{align}
\inf_T (\mathrm{ or }\sup_T)  ~\mathcal{M}(T), \quad \mathrm{subject~to}~ t_a\left(T\right)\geq \gamma, \label{STmtrix}
\end{align}
where $\mathcal{M}(T)$ indicates some detection performance achieved by stopping time $T$, and $\gamma$ is a pre-designed value that constitutes a lower bound on ARL2FA.

In general, different instances of $\mathcal{M}(\cdot)$ in \eqref{STmtrix} requires different test statistics $Z_t$ and detection thresholds $\eta$ in \eqref{GeneralizedStoppingTimeForm}.
In the following, we introduce three widely discussed non-Bayesian SCPD procedures:
1) the Shewhart test,
2) the CUSUM test, and
3) the SR test.
For clarity, in the following, we use $\{\eta_s,T_s, S_t\}$, $\{\eta_c,T_c,C_t\}$, and $\{\eta_r,T_r,R_t\}$ to denote detection thresholds, stopping times, and test statistics in Shewhart, CUSUM, and SR tests, respectively.
For notational simplicity in the following, we use $\mathbb{E}_{k}\left(\cdot\right)$ and $\mathbb{P}_{k}\left(\cdot\right)$ to denote $\mathbb{E}_{k}\left(\cdot | \nu = k \right)$ and $\mathbb{P}_{k}\left(\cdot | \nu = k \right)$, respectively, and specifically $\mathbb{E}_{\infty}\left(\cdot\right)$ and $\mathbb{P}_{\infty}\left(\cdot\right)$ to denote $\mathbb{E}\left(\cdot | \nu = \infty \right)$ and $\mathbb{P}\left(\cdot | \nu = \infty \right)$, respectively.

\subsubsection{Shewhart test}
\label{sssh}
In the Shewhart test, the test statistic at time $t$ is given by
\begin{align}
Z_t = S_t \triangleq \Lambda(y_{t}) , \label{ShewhartTestStatistic}
\end{align}
where $\Lambda(x)\triangleq\frac{f_{1}(x)}{f_{0}(x)}$ is the likelihood ratio with $f_{1}(x)$ and $f_0(x)$ being the pre-change and post-change PDFs, respectively. In our case, according to \eqref{ReceivedSignalSequenceModified},
we have $\Lambda(x) = (1/(q+1)) \mathrm{e}^{q|x|^2/(1+q) }$.
Obviously, the test statistic in the Shewhart test only depends on the instantaneous sample observation and is independent of  past observations.
Note that the Shewhart test is optimal under several performance metrics.
In particular, as discussed in \cite{M.Pollak2013,G.V.Moustakides2014}
if $\eta_s$ satisfies $\mathbb{P}_{\infty}\{\Lambda(y_1)\geq \eta_s\}=1/\gamma$, then subject to the ARL2FA constraint in \eqref{STmtrix},
the Shewhart test attains the highest probability to immediately and successfully recognize the change-point in the worst case of $\nu$.
We will discuss how to obtain the value of $\eta_s$ that satisfies $\mathbb{P}_{\infty}\{\Lambda(y_1)\geq \eta_s\}=1/\gamma$ in Section \ref{QLDerivationSH}.

\subsubsection{CUSUM test}

\label{Sec:CUSUMIntroSec}
The CUSUM test was first proposed in \cite{E.S.Page1954}.
In the CUSUM test, at time $t$, the test statistic is
\begin{align}
Z_t = C_t\triangleq\max_{1 \leq k \leq t } \sum_{j=k+1}^t \ln \Lambda(y_{j}),
\label{CUSUMTestStatistic}
\end{align}
where it is understood that $\sum_{j=t+1}^t \left(\cdot\right)= 0$.
Note that $C_t$ can also be written in a recursive form, i.e.,  $C_t = \max\{0,C_{t-1} + \ln \Lambda(y_{t})\}$ with $C_0 = 0$.
Under the condition that the change is persistent, namely $L=\infty$ in \eqref{ReceivedSignalSequenceModified}, it was shown in  \cite{G.V.Moustakides1986} that, if $\eta_c$ satisfies $t_a\left(T_c\right) = \gamma$, then the CUSUM test minimizes the average detection delay in the worst case of the change-point $\nu$ and the pre-change sample observations $\{y_1,y_2,\cdots,y_{\nu-1}\}$.
Discussions on how to obtain the value of $\eta_c$ that satisfies $t_a\left(T_c\right) = \gamma$ will be presented in Section \ref{ThresholdinCUSUMandSR}.

\subsubsection{SR test}
\label{sssr}
In the SR test, the test statistic at time $t$ is
\begin{align}
Z_t = R_t\triangleq \sum_{k=1}^t \prod_{j=k}^t \Lambda(y_{j}).\label{SRTestOutPut}
\end{align}
Note that similar to the case in the CUSUM test, $R_t$ can also be written in a recursive form $R_t = \left(1 + R_{t-1} \right)\Lambda(y_{t})$ with $R_0 = 0$.
Under the condition that the change is persistent, namely $L=\infty$ in \eqref{ReceivedSignalSequenceModified}, it was proved in \cite{M.Pollak2008} that for any $\gamma \geq 1$,  the SR test minimizes the relative integral average detection delay when the detection threshold $\eta_r$ satisfies $t_a\left(T_r\right) = \gamma$.
It was also pointed out in \cite{M.Pollak2008} that the SR test is optimal in the sense that it minimizes the average detection delay under the condition that $\nu$ is sufficiently large and the detection procedure is cyclic and preceded by a stationary flow of false alarms. We will show how to obtain the value of $\eta_r$ that satisfies $t_a\left(T_r\right) = \gamma$ in Section \ref{ThresholdinCUSUMandSR}.

In this paper, we assume that Willie adopts one of the above three tests to detect LPD communication between Alice and Bob, and we evaluate the performance of LPD communication under these tests.

%

\begin{remark}
\label{Remark:DifferentKnowledge}
To enable the CUSUM or SR test, Willie needs to know the transmit power of Alice, i.e., $q$. Otherwise, he is not able to compute the test statistics $C_t$ and $R_t$ according to \eqref{CUSUMTestStatistic} and \eqref{SRTestOutPut}.
For the Shewhart test, it can be verified that the detection procedure is equivalent to sequentially comparing the received signal power to a pre-designed threshold, and therefore, the value of $q$ is not required.
\end{remark}
\subsection{Problem statement of LPD communication}
\label{Section:LPDProState}
\begin{figure}[t]
  \centering
  \includegraphics[width=3.5 in]{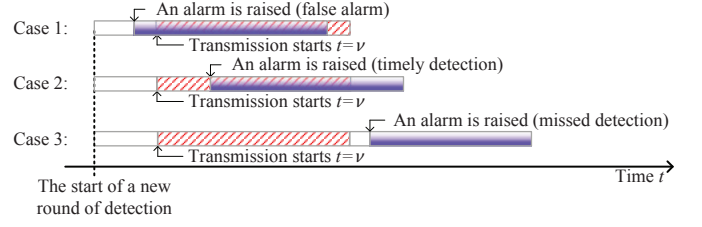}
  \caption{\small Three cases regarding to the start time of Alice's transmission, where the blue shaded regions represent that Willie is dealing with the alarm raised by his detector, and the regions filled with red slashes represents that Alice is transmitting. }\label{Model3Cases}
  \vspace{-1mm}
\end{figure}

Consider that Alice starts to transmit to Bob at time $t = \nu+1$.
On one hand, practical SCPD usually stops within finite time with probability one, i.e., $\mathbb{P}\{T< \infty\} = 1$ \cite{H.V.Poor2009,A.G.Tartakovsky2013}. Consequently,
as long as Alice transmits for a sufficiently long time with a strictly positive power level, Alice's transmission will always be detected by Willie at some time after $\nu$.
On the other hand, there exists an inevitable detection delay in practical SCPD, i.e.,
there is an interval between the time when the transmission is successfully detected  and the time when it truly starts.
Therefore, an effective covert transmission scheme only transmits for a short period of time, i.e., $L<\infty$.
Specifically, the goal of Alice is to transmit as much information as possible to Bob during $L$ channel usages such that her probability of being detected falls below a threshold.
As Willie's detector exhibits an inevitable detection delay, if
$L$ is significantly smaller than the detection delay, Willie is unlikely to be aware of Alice's transmission.
Based on the detection process in Fig. \ref{SCPDModel},  there are three different cases regarding to the alarm raising time, which are illustrated in Fig. \ref{Model3Cases}.
In Case 1, an alarm is raised before the transmission starts,which is a \emph{false alarm}.
In Case 2, Alice's transmission is detected, and an alarm is raised during Alice's transmission. We refer to such an event as \emph{timely detection}.
In Case 3, Alice's transmission is detected but the alarm is raised after Alice has already accomplished the transmission, and such an event is referred to as \emph{missed detection}.

Note that each time   an alarm is raised, Willie will take   action  to handle it. As a result, a false alarm can impact on the communication between Alice and Bob, as indicated in Case 1 of Fig. \ref{Model3Cases}.
However, in many practical scenarios, such a case can be avoided by the careful design of the communication protocol between Alice and Bob. An example is presented below to demonstrate this.

\vspace{0.1cm}
\noindent
\textbf{Example:}
\textit{Consider a scenario where each time when Willie's detector raises an alarm, he jams the wireless channel for a period of time to interrupt the communication between Alice and Bob. In this way, if Alice is indeed transmitting, then Bob will not be able to decode the message transmitted by Alice as the channel is jammed. Even if the alarm is false, this strategy will not cost too much for Willie as long as the frequency of false alarm is designed to be sufficiently low. From Alice's and Bob's perspective, if a false alarm is raised before their communication and triggers Willie's jamming, then both Alice and Bob will receive the jamming signal. As a result, they can deliberately delay the start time of their communication to avoid being jammed.}
\vspace{0.1cm}

Motivated by the example above, Alice and Bob can effectively avoid the occurrence of Case 1 if they can sense Willie's response to the alarms.
In this paper, we assume that Alice and Bob are capable of doing so.
As a result, for a given data packet to be transmitted,
Alice need only consider the events of timely detection and missed detection as illustrated in Fig. \ref{Model3Cases}
\footnote{
This does not mean that the false alarms raised by Willie's detector have no impact on the communication between Alice and Bob.
In fact, each alarm raised by Willie's detector precludes Alice from transmitting to Bob for a period of time (depending on how Willie handles the alarms).
Particularly, the false alarm raised right before the start of Alice's transmission possibly forces Alice to delay her transmission, the effect of which can be significant in application scenarios with low latency requirement.
Note that such an effect is out the scope of this paper but constitutes a future research issue.
}.
Under the condition that Case 1 does not occur, we observe that on the one hand, timely detection or missed detection occurs with probability one due to the fact that Willie's detection process terminates within finite time, i.e, $\mathbb{P}\{T < \infty\} = 1$ \cite{H.V.Poor2009,A.G.Tartakovsky2013}; on the other hand, in many application scenarios, Willie's detection process makes sense only when timely detection occurs, an example of which is presented above.
Based on this viewpoint,
we define \emph{covert probability}, denoted by $Q_L(q)$, as the probability that Willie is not alerted during the transmission, i.e., the probability of a missed detection,
\begin{align}
&\quad \mathcal{Q}_{L}\left(q\right) \triangleq
\mathbb{P}\left\{ T>\nu+L | T > \nu\right\} \nonumber \\
& =
\mathbb{P}\left\{Z_{\nu + 1} < \eta ,Z_{\nu + 2} < \eta,\cdots, Z_{\nu+L} < \eta  | T > \nu\right\}. \label{DefinitionFailDetection}
\end{align}
And based on above discussions, Alice and Bob expect a sufficiently high covert probability in order for the communication between them to be covert.

Note that for a given $(q,L)$ pair, different SCPD algorithms generate different covert probabilities, therefore, we use $\mathcal{Q}_L^{\mathrm{SH}}(q)$, $\mathcal{Q}_L^{\mathrm{CU}}(q)$, and $\mathcal{Q}_L^{\mathrm{SR}}(q)$ to denote the covert probabilities in the Shewhart, the CUSUM, and the SR tests, respectively.

In this paper, we use the following utility function to characterize  LPD communication performance,
\begin{align}
\mathcal{I}\left(q,L\right) \triangleq  L\ln\left( 1 + q/\sigma_B^2 \right), \label{UtilityFunction}
\end{align}
where $\sigma_B^2$ is the noise power at Bob. As we assume a finite length of transmission time, the utility function in \eqref{UtilityFunction} serves as an upper bound on the total amount of information that can be conveyed from Alice to Bob covertly.

Based on the above considerations, Alice's problem is,
\begin{align}
\label{CovertCapacity}
\max_{q,L} \quad \mathcal{I}\left(q,L\right),  \quad \mathrm{s.t.} \quad \mathcal{Q}_L(q) \geq \theta,
\end{align}
where $\theta$ ($0 < \theta < 1$) is a pre-designed threshold and constraint $\mathcal{Q}_L(q) \geq \theta$ means that the probability that Willie is not alerted during the transmission is lower bounded by $\theta$.

In the following sections of this paper, we first analyze the covert probabilities under the three sequential tests introduced in Section \ref{IntroductionST}, and then we numerically evaluate LPD communication performance by solving the maximization problem in \eqref{CovertCapacity}.

\begin{remark}
\label{PriorKnowledgeToSolveProblem}
To solve \eqref{CovertCapacity}, Alice needs to know the noise variance at Willie's receiver $\sigma_W^2$ and the ARL2FA of Willie's detector $\gamma$.
In practice, it is possible for Alice to estimate these parameters.
For example, the noise power is usually influenced by the surrounding electromagnetic environment. If Alice is located close to Willie such that they are in the same electromagnetic environment, then Alice is able to roughly estimate $\sigma_W^2$.
As for the ARL2FA of Willie's detector, we note that each alarm raised by Willie's detector is accompanied by Willie's response to deal with it.
One example is presented above, wherein Willie jams the channel each time an alarm is raised in order to disrupt the communication between Alice and Bob. Therefore, it is possible for Alice to estimate the ARL2FA by recording the occurrence time of Willie's responses.
Based on these discussions, we assume that Alice knows the values of $\sigma_W^2$ and $\gamma$ in this paper. In practice, Alice's knowledge about $\sigma_W^2$ and $\gamma$ can be inaccurate due to the estimation errors.
Analyzing LPD communication performance when Alice only has inaccurate knowledge about $\sigma_W^2$ and $\gamma$ is more challenging and left for future research.
\end{remark}

\begin{remark}
In this paper, the ARL2FA of Willie's detector, $\gamma$, is assumed to be a fixed constant.
In practice, the value of $\gamma$, which controls the false alarm frequency,
is highly related to the cost that Willie pays due to   alarms raised by his detector (for instance, the energy consumption required to jam the channel in the example presented above).
The decrease in $\gamma$ causes the increase in the cost because Willie needs to respond to the false alarms more frequently if LPD communication does not occur between Alice and Bob.
Though such a cost can be reduced by increasing $\gamma$, this generally results in a longer detection delay, which can be exploited by Alice and Bob to achieve a higher covert throughput.
As a result, instead of fixing $\gamma$ as a constant, Willie might optimize $\gamma$ based on Alice's transmission scheme to reach a trade-off between the cost due to the false alarms and the covert throughput that can be achieved by Alice and Bob.
From this viewpoint, the problem of LPD communication can be modeled as a game played by Alice and Willie,
wherein Willie carefully designs his detector (including the value of $\gamma$) to minimize the overall loss,
on top of which Alice selects a transmission scheme that achieves the best LPD communication performance.
Solving the problem of LPD communication from the perspective of game theory constitutes an interesting research problem.
\end{remark}

\section{Covert probability analysis}
In this section, we assume that Willie adopts one of the aforementioned tests to discover LPD communication. We investigate the probability that Willie fails to detect the transmission from Alice to Bob during the first $L$ channel usages.

\subsection{Shewhart test}
\label{QLDerivationSH}
With the Shewhart test, the following theorem characterizes the probability that Willie does not successfully detect the ongoing transmission during the first $L$ channel usages.

\begin{theorem}
For a given threshold $\eta_s$ in the Shewhart test, $\mathcal{Q}_L^{\mathrm{SH}}(q)$ is given by
\begin{align}
\mathcal{Q}_L^{\mathrm{SH}}(q) = \left(1 - \mathrm{e}^{-\frac{1}{q}\ln\left(\left(1+q\right)\eta_s\right)}
\right)^L. \label{ShewarttestQL}
\end{align}
\end{theorem}
\begin{IEEEproof}
According to  \eqref{DefinitionFailDetection}, $\mathcal{Q}_L^{\mathrm{SH}}$ can be written as
\begin{align}
&\quad \mathcal{Q}_L^{\mathrm{SH}}(q)\overset{(a)}{=}
\mathbb{P}\left\{S_{\nu+1} < \eta_s, \cdots,S_{\nu+L} < \eta_s\right\}\nonumber \\
&= \mathbb{P}\left\{
\frac{1}{1+q}\mathrm{e}^{\frac{q}{1+q}\left|y_{\nu+1}\right|^2} < \eta_s,
\cdots,
\frac{1}{1+q}\mathrm{e}^{\frac{q}{1+q}\left|y_{\nu+L}\right|^2} < \eta_s\right\} \nonumber \\
&
\overset{(b)}{=} \prod_{k=1}^L\mathbb{P}\left\{\frac{\mathrm{e}^{\frac{q}{1+q}
\left|y_{\nu+k}\right|^2}}{1+q} < \eta_s \right\} \overset{(c)}{=} \left(1 - \mathrm{e}^{-\frac{\eta_s^{\prime}}{1+q}}\right)^L
\end{align}
where $\eta_s^{\prime}\triangleq\frac{1+q}{q}\ln\left((1+q)\eta_s\right)$, $(a)$ follows from the fact that $\{S_{\nu+k}:1\leq k\leq L\}$ is independent of $\{S_{k}:1\leq k\leq \nu\}$,
$(b)$ follows from the fact that $\{y_{\nu+k}:1\leq k\leq L \}$ are i.i.d. random variables, and
$(c)$ is because $\left|y_{\nu+k}\right|^2\sim\mathcal{E}\left(1+q\right)$ for $k>1$.
\end{IEEEproof}

Note that the optimal Shewhart test requires that $\mathbb{P}_{\infty}\left\{\Lambda(y_{1}) \geq \eta_s\right\} = \frac{1}{\gamma}$ \cite{M.Pollak2013,G.V.Moustakides2014}, i.e.,
\begin{align}
\mathbb{P}_{\infty}\left\{\Lambda(y_{1}) \geq \eta_s\right\} & =
\mathbb{P}_{\infty}\left\{ \left|y_{1}\right|^2 > \eta_s^{\prime} \right\} \overset{(a)}{=} \mathrm{e}^{-\eta_s^{\prime}} = \frac{1}{\gamma},\label{ShewarttestBasedOntheFalseAlarm} \\
\Rightarrow~~\eta_s^{\prime} &= \ln (\gamma),
\end{align}
where $(a)$ follows from the fact $\left|y_{1}\right|^2\sim \mathcal{E}\left(1\right)$ under the condition $\nu = \infty$. Therefore, by inserting \eqref{ShewarttestBasedOntheFalseAlarm} into \eqref{ShewarttestQL}, $\mathcal{Q}_L^{\mathrm{SH}}(q)$ can be further simplified as
\begin{align}
\mathcal{Q}_L^{\mathrm{SH}}(q) = \left(1 - \left( 1/\gamma \right)^{\frac{1}{1+q}} \right)^L.\nonumber
\end{align}

\subsection{CUSUM test}
In this subsection, we derive the covert probability under the condition that Willie performs the CUSUM test.
According to \eqref{CUSUMTestStatistic}, the CUSUM test statistic can be written as
$C_{t} = \max\{0, C_{t-1} + \frac{q}{1+q}\left|y_t\right|^2 - \ln\left(1 + q\right)\}$ with $C_0 = 0$.
To derive the covert probability, for notational simplicity, we scale the CUSUM test statistic, $C_{t}$, as
$\hat{C}_{t} \triangleq \frac{1+q}{q}C_{t}$ for $t\geq 1$, and therefore, we have
\begin{subequations}
\label{ScaledCUSUMtestSta}
\begin{align}
\hat{C}_{t} &= \max\left\{0, \hat{C}_{t-1} + X_t - \omega \right\},\quad \hat{C}_0 = 0,\quad t\geq 1,\\
T_c &= \inf\{t: t \geq 1, \ \hat{C}_{t}\geq \hat{\eta}_c \},
\end{align}
\end{subequations}
where $X_t\triangleq \left|y_t\right|^2$,
$\omega\triangleq\frac{1+q}{q}\ln\left(1 + q\right)$, and
$\hat{\eta}_c\triangleq \frac{1+q}{q}\eta_c$ is the scaled detection threshold.

The following lemma presents the conditional cumulative distribution function (CDF) of $\hat{C}_{t+1}$, which will be useful in deriving $\mathcal{Q}_{L}^{\mathrm{CU}}(q)$.
\begin{lemma}
\label{CUSUMSTATISTICCONDITIONCDF}
Given $\hat{C}_{t} = u$, the conditional CDF of $\hat{C}_{t+1}$, denoted by $\mathcal{P}_{t+1|t}^{\mathrm{CU}} (x|u)$, is
\begin{align}\label{ConditionCnp1GivenCn}
& \quad  \mathcal{P}_{t+1|t}^{\mathrm{CU}} (x|u)\triangleq
\mathbb{P}\left\{\hat{C}_{t+1} \leq x | \hat{C}_{t} = u\right\} \nonumber\\
&=\left\{
\begin{aligned}
\mathcal{P}_{0}^{\mathrm{CU}} (x|u) &\triangleq  \left(1 - \mathrm{e}^{- \frac{\left(\omega - u + x\right)}{1+q}} \right) \\
&\quad \times \mathbb{I}\left\{x \geq \max\{ u - \omega , 0 \} \right\},  \text{if }t\geq \nu,\\
\mathcal{P}_{\infty}^{\mathrm{CU}} (x|u) &\triangleq \left(1 -  \mathrm{e}^{ -\left(\omega - u + x\right) } \right) \\
&\quad \times \mathbb{I}\left\{x \geq \max\{ u - \omega , 0 \} \right\},  \text{if }t< \nu.
\end{aligned}
\right.
\end{align}
\end{lemma}
\begin{IEEEproof}
By assumption $X_t\sim \mathcal{E}(1)$ when $t\leq\nu$ and $X_t\sim \mathcal{E}(1+q)$ when $t>\nu$.
\end{IEEEproof}

Now, we derive $\mathcal{Q}_L^{\mathrm{CU}}(q)$. To do this, for $n \geq 1$, define
\begin{align}
\mathcal{Q}_n^{\mathrm{CU}}(x;q)
& = \mathbb{P}\big\{\hat{C}_{\nu+1} \leq \hat{\eta}_c,\hat{C}_{\nu+2} \leq \hat{\eta}_c,\nonumber \\
&\quad\quad\quad \cdots,\hat{C}_{\nu+n} \leq \hat{\eta}_c | \hat{C}_{\nu} = x, T_c>\nu\big\}, \label{ConditionalCUSUMQL}
\end{align}
where $0\leq x < \hat{\eta}_c$.
Obviously, by letting $n=L$ in \eqref{ConditionalCUSUMQL}, $\mathcal{Q}_L^{\mathrm{CU}}(x;q)$ can be viewed as the conditional covert probability conditioned on $\hat{C}_{\nu} = x$. By the law of total probability, we have $\mathcal{Q}_L^{\mathrm{CU}}(q)= \mathbb{E}_{\hat{C}_{\nu}}\left\{\mathcal{Q}_L^{\mathrm{CU}}(\hat{C}_{\nu};q)\right\}$.
Therefore, to obtain $\mathcal{Q}_L^{\mathrm{CU}}(q)$, in the following, we first derive $\mathcal{Q}_n^{\mathrm{CU}}(x;q)$ and then investigate the distribution of $\hat{C}_{\nu}$.

Before presenting an expression for $\mathcal{Q}_n^{\mathrm{CU}}(x;q)$,
we introduce some constants that do not depend on $x$, which are useful in presenting $\mathcal{Q}_n^{\mathrm{CU}}(x;q)$.
We denote $M$ as the smallest integer that satisfies $\hat{\eta}_c \leq M \omega$, i.e.,  $M\triangleq\left\lceil\frac{\hat{\eta}_c}{\omega}\right\rceil$.
Define $\mathcal{V}_1\triangleq\{V_{1,1,1},V_{1,1,2}\}$ with $V_{1,1,1}=1$ and $V_{1,1,2}= -\mathrm{e}^{-\frac{\hat{\eta}_c+\omega}{1+q}}$.
For $n\geq 1$, define $\mathcal{V}_{n+1}\triangleq\{V_{n+1,i,j}: 1\leq i \leq (n+1)\wedge M, 1\leq j\leq i + 1\}$ with
\begin{align}
\begin{aligned}
 V_{n+1,1,1} &= V_{n,1,1} + V_{n,1,2},   \\
 V_{n+1,i,1} &= V_{n,i-1,1}, \quad  \text{if } 2 \leq i \leq \left(n+1\right)\wedge M,   \\
 V_{n+1,i,j} &= \alpha V_{n,i-1,j-1}, \text{if }
 \left\{ \begin{aligned}& 2 \leq i \leq \left(n+1\right)\wedge M, \\ &3 \leq j \leq i+1 \end{aligned}\right\},   \\
 V_{n+1,i,2} &=
 \left\{\begin{aligned}
 &-V_{n+1,i,1}\beta_i  \\
 & \quad\quad  + \alpha \sum\nolimits_{k=i}^{n\wedge M} \Psi_{n,k},\text{if } 1\leq i \leq n\wedge M,   \\
 &-V_{n+1,n+1,1} \beta_1 \underline{\beta}+ \alpha \sum\nolimits_{k=1}^n \vartheta_{0,k-1,k} \\
 &\quad\quad  \times  V_{n,n,k+1},\text{if } \left\{\begin{aligned}&i = n+1,\\ &n < M\end{aligned}\right\},
 \end{aligned}\right.
\end{aligned}
\end{align}
where
$\alpha \triangleq \frac{1}{1+q}\mathrm{e}^{-\frac{\omega}{1+q}}$,
$\beta_i \triangleq \mathrm{e}^{-\frac{i\omega}{1+q}}$,
$\underline{\beta} \triangleq \mathrm{e}^{-\frac{\hat{\eta}_c}{1+q}}$,
$\vartheta_{a,b,l} \triangleq \frac{(a\omega)^l - (b\omega - \hat{\eta}_c)^l}{l!}$ for any integer-valued $a$, $b$, and $l$, and
$\{\Psi_{n,k}: n\geq 1,1\leq k\leq  n \wedge M\}$ is given by
\begin{align}
\Psi_{n,k} =
\left\{
\begin{aligned}
&(1+q)(\beta_{k-1} - \beta_k)V_{n,k,1} \\
&\quad  + \sum\nolimits_{l=1}^k V_{n,k,l+1} \frac{\omega^l}{l!}, \text{if }1\leq k < n \wedge M, \\
&(1+q)\left(\beta_{n-1} - \underline{\beta}\right)V_{n,n,1} \\
&\quad + \sum\nolimits_{l=1}^n V_{n,n,l+1} \vartheta_{l-n,l-1,l}, \text{if }k = n , n \leq M, \\
&(1+q)\left(\beta_{M-1} - \underline{\beta}\right)V_{n,M,1}\\
&\quad + \sum\nolimits_{l=1}^M V_{n,M,l+1} \vartheta_{1,M,l}, \text{if }k = M , n > M. \\
\end{aligned}
\right.
\end{align}
Note that $\mathcal{V}_{n+1}$ only depends on
$\mathcal{V}_{n}$, and can be recursively calculated from $\mathcal{V}_1$.

Based on $\mathcal{V}_n$ ($n\geq 1$) defined above, an exact theoretical expression of $\mathcal{Q}_n^{\mathrm{CU}}(x;q)$ is given in the following theorem.
\begin{theorem}
\label{Therorem:CUSUMQL}
For $n\geq 1$ and $x\in [0,\hat{\eta}_c)$, $\mathcal{Q}_n^{\mathrm{CU}}(x;q)$ is given by \eqref{QNXQ},
\begin{figure*}[t]
\begin{align}
\label{QNXQ}
\mathcal{Q}_n^{\mathrm{CU}}(x;q)= \left\{
\begin{aligned}
& V_{n,i(x),1} + \sum_{j=1}^{i(x)} V_{n,i(x),j+1} \frac{\left(i(x)\omega - x\right)^{j-1}}{(j-1)!}\mathrm{e}^{\frac{x}{1 + q}} , &&\text{ if } x \in [0 , (n-1)\omega \wedge \hat{\eta}_c ),\\
& V_{n,n,1} + \sum_{j=1}^{n} V_{n,n,j+1} \frac{\left((j-1)\omega - x\right)^{j-1}}{(j-1)!}\mathrm{e}^{\frac{x}{1 + q}},&&\text{ if } x \in [(n-1)\omega , \hat{\eta}_c).
\end{aligned}
\right.
\end{align}
\noindent\rule[0.25\baselineskip]{\textwidth}{1pt}
\vspace{-10mm}
\end{figure*}
where $i(x)$ is the unique integer that satisfies $(i(x)-1)\omega \leq x < i(x)\omega$.
\end{theorem}
\begin{IEEEproof}
$\mathcal{Q}_{1}^{\mathrm{CU}}(x;q)$ is obtained by using Lemma 1.
The derivation of $\mathcal{Q}_{n}^{\mathrm{CU}}(x;q)$ with $n\geq 2$ is in Appendix \ref{Appendix:A}.
\end{IEEEproof}

Theorem \ref{Therorem:CUSUMQL} provides us a way to evaluate the conditional covert probability, i.e., $\mathcal{Q}_{L}^{CU}(x;q)$.
Now, we derive the distribution of $\hat{C}_{\nu}$.
We denote the conditional CDF of $\hat{C}_{\nu}$ as
\begin{align}
\mathcal{G}_{\nu}^{\mathrm{CU}}(x;q)
&\triangleq \mathbb{P}\left\{\hat{C}_{\nu} \leq x | T_c>\nu \right\},\quad 0\leq x < \hat{\eta}_c.
 \label{ConditionalCnu}
\end{align}
To obtain $\mathcal{G}_{\nu}^{\mathrm{CU}}(x;q)$, for $n$ ($1\leq n \leq \nu$), we define
\begin{align}
\begin{aligned}
\mathcal{G}_n^{\mathrm{CU}}(x;q) &\triangleq \mathbb{P}\{\hat{C}_{n} \leq x | T_c > n  \},\\
\tilde{\mathcal{G}}_n^{\mathrm{CU}}(x;q) &\triangleq \mathbb{P}\{\hat{C}_{n} \leq x | T_c > n-1 \}.
\end{aligned}
\end{align}
It is easy to see that $\mathcal{G}_n^{\mathrm{CU}}(x;q) = \tilde{\mathcal{G}}_n^{\mathrm{CU}}(x;q)/\tilde{\mathcal{G}}_n^{\mathrm{CU}}(\hat{\eta}_c;q) $ for $0\leq x < \hat{\eta}_c$.

Before presenting expressions for $\tilde{\mathcal{G}}_n^{\mathrm{CU}}(x;q)$ and $\mathcal{G}_n^{\mathrm{CU}}(x;q)$, we specify some constants that are useful in our derivations. We define $\mathcal{A}_1\triangleq\{\tilde{A}_{1,1,1},\tilde{A}_{1,1,2}\}$, where $\tilde{A}_{1,1,1} = 1$ and $\tilde{A}_{1,1,2}=\mathrm{e}^{-\omega}$.
For $n \geq 1$, we define
$\mathcal{A}_{n+1}\triangleq\{\tilde{A}_{n+1,i,j}: 1\leq i \leq \left(n+1\right) \wedge M, 1\leq j\leq i+1\}$ with
$\tilde{A}_{n+1,1,1} = 1$ and
\begin{align}
\begin{aligned}
&\tilde{A}_{n+1,i,1} = A_{n,i-1,1}, \quad \quad \quad\  \text{if } 2\leq i \leq (n+1)\wedge M, \\
&\tilde{A}_{n+1,i,j} = A_{n,i-1,j-1}\mathrm{e}^{-\omega},\quad
\text{if }\left\{\begin{aligned} &2\leq i \leq (n+1)\wedge M,\\ &3\leq j \leq i+1\end{aligned}\right\}, \\
&\tilde{A}_{n+1,i,2} =
\left\{\begin{aligned}
&\tilde{A}_{n+1,i,1} \delta_i - \mathrm{e}^{-\omega}\sum\nolimits_{l=1}^{i-1} A_{n,i-1,l+1} \varsigma_{1,0,l} \\
&\quad\quad - \mathrm{e}^{-\omega}\sum\nolimits_{k=i}^{n\wedge M}  \Upsilon_{n,k},
~\text{if }1 \leq i \leq n\wedge M, \\
& - \mathrm{e}^{-\omega}\left(\sum\nolimits_{l=1}^{n} A_{n,n,l+1}\varsigma_{0,l-1,l}\right)\\
&\quad\quad + \mathrm{e}^{ -\omega} \tilde{A}_{n+1,n+1,1},
~\text{if }\left\{\begin{aligned}&i = n+1,\\& n < M\end{aligned}\right\}, \\
\end{aligned}\right. \\
& A_{n+1,i,j}=  \tilde{A}_{n+1,i,j} \Big  /\left( \tilde{A}_{n+1,1,1} - \tilde{A}_{n+1,1,2}\mathrm{e}^{- \hat{\eta}_c} \right),
\end{aligned}\nonumber
\end{align}
where $\delta_i \triangleq \mathrm{e}^{\hat{\eta}_c - i\omega}$, $\varsigma_{a,b,l}\triangleq\frac{(a\hat{\eta}_c + b\omega)^l}{l!}$ for any integer $a$, $b$, and $l$, and $\{\Upsilon_{n,k}:n \geq 1,1\leq k \leq n\wedge M\}$ is given by
\begin{align}
\Upsilon_{n,k} =
\left\{\begin{aligned}
& A_{n,k,1}\left(\delta_{i-1} - \delta_{i}\right)- \sum\nolimits_{l=1}^{k} A_{n,k,l+1}\\
&\quad\quad\quad \times\left(\varsigma_{1,1,l} - \varsigma_{1,0,l}\right),\quad\quad\text{if }1\leq k < n \wedge M,\\
& A_{n,n,1}\left(\delta_{n-1} - 1\right) - \sum\nolimits_{l=1}^{k} A_{n,n,l+1}\\
&\quad\quad\quad \times\left(\varsigma_{1,l-n,l} - \varsigma_{0,l-1,l}\right),~\text{if }k = n, n \leq M,\\
& A_{n,M,1}\left(\delta_{M-1} - 1\right) - \sum\nolimits_{l=1}^{M} A_{n,M,l+1}\\
&\quad\quad\quad \times\left(\varsigma_{1,1,l} - \varsigma_{0,M,l}\right),\quad\quad\text{if }k = M, n > M. \\
\end{aligned}\right.
\end{align}
Note that $\mathcal{A}_{n+1}$ only depends on
$\mathcal{A}_{n}$ and can be recursively calculated from $\mathcal{A}_1$.

Based on $\mathcal{A}_n$ ($n\geq 1$) defined above, we present $\tilde{\mathcal{G}}_n^{\mathrm{CU}}(x;q)$ and $\mathcal{G}_n^{\mathrm{CU}}(x;q)$ in the following theorem.
\begin{theorem}
\label{Theorem:StationaryCUSUMdistribution}
For $n\geq1$ and $x\in [0,\hat{\eta}_c)$, $\tilde{\mathcal{G}}_n^{\mathrm{CU}}(x;q)$ is given by \eqref{TGNxq}
\begin{figure*}[t]
\begin{align}
\tilde{\mathcal{G}}_n^{\mathrm{CU}}(x;q)
= \left\{
\begin{aligned}
& \tilde{A}_{n,\hat{i}(x),1} - \sum_{j=1}^{\hat{i}(x)} \tilde{A}_{n,\hat{i}(x),j+1}\frac{(x+\hat{i}(x)\omega)^{j-1}}{(j-1)!}\mathrm{e}^{- x},&& \text{ if } 0\vee \left(\hat{\eta}_c - (n-1)\omega\right) \leq x < \hat{\eta}_c,\\
& \tilde{A}_{n,n,1} - \sum_{j=1}^{n} \tilde{A}_{n,n,j+1}\frac{(x+(j-1)\omega)^{j-1}}{(j-1)!}\mathrm{e}^{- x},&& \text{ if } 0 \leq x < \hat{\eta}_c - (n-1)\omega.\\
\end{aligned}
\right.
\label{TGNxq}
\end{align}
\noindent\rule[0.25\baselineskip]{\textwidth}{1pt}
\vspace{-7mm}
\end{figure*}
where $\hat{i}(x)$ is the unique integer that satisfies $\hat{\eta}_c - \hat{i}(x) \omega \leq x < \hat{\eta}_c - (\hat{i}(x)-1) \omega $.
Based on $\tilde{\mathcal{G}}_n^{\mathrm{CU}}(x;q)$, $\mathcal{G}_{\nu}^{\mathrm{CU}}(x;q)$ can be written as
$\mathcal{G}_{\nu}^{\mathrm{CU}}(x;q) =  \tilde{\mathcal{G}}_{\nu}^{\mathrm{CU}}(x;q)\Big/\left(\tilde{A}_{\nu,1,1} - \tilde{A}_{\nu,1,2}\mathrm{e}^{-\hat{\eta}_c} \right)$.
\end{theorem}
\begin{IEEEproof}
The proof is found in Appendix \ref{ProofPRIORCnu}.
\end{IEEEproof}

With $\mathcal{Q}_{L}^{\mathrm{CU}}(x;q)$ and $\mathcal{G}_{\nu}^{\mathrm{CU}}(x;q)$ derived in Theorems \ref{Therorem:CUSUMQL} and  \ref{Theorem:StationaryCUSUMdistribution}, respectively, $\mathcal{Q}_{L}^{\mathrm{CU}}(q)$ can be written as
$\mathcal{Q}_L^{\mathrm{CU}}(q) =
\int_{0}^{\hat{\eta}_c} \mathcal{Q}_L^{\mathrm{CU}}(x;q) \mathrm{d} \mathcal{G}_{\nu}^{\mathrm{CU}}(x;q)$.
We note that, in general, $\mathcal{Q}_{L}^{\mathrm{CU}}(q)$ depends on the value of $\nu$.

\subsection{SR test}
\label{CovertAnalysisSRTest}
According to \eqref{SRTestOutPut}, the SR test statistic at time $t$, i.e., $R_t$, is
\begin{align}
R_t &= \left(1 + R_{t-1}\right)\Lambda\left(y_t\right) =
\frac{\left(1 + R_{t-1}\right)}{1+q}\mathrm{e}^{\frac{q}{1+q}\left|y_t\right|^2},
\label{RecursiveSRStatistic}
\end{align}
with $R_0 = 0$.
Therefore, $R_t\geq \frac{1+R_{t-1}}{1+q}\geq \frac{1+R_{t-1}}{(1+q)^2} + \frac{1}{(1+q)}\geq\cdots\geq \frac{1}{q}\left(1 - \frac{1}{(1+q)^t}\right)$. Regarding the detection threshold $\eta_r$, for ease of discussion, we only consider the case $\eta_r \geq \frac{1}{q}$. This is because when $\eta_r < \frac{1}{q}$, the SR detector always raises a false alarm before the transmission begins provided $\nu>\frac{- \ln\left(q (\frac{1}{q}-\eta_r)\right)}{\ln(1+q)}$, which has also been pointed out in \cite{W.Du2015}.
Nevertheless, the method used in this part to evaluate $\mathcal{Q}_L^{SR}(q)$ can also be extended to the case with $\eta_r < \frac{1}{q}$.

Based on \eqref{RecursiveSRStatistic}, the sequence $\{R_t:t\geq1\}$ forms a Markov chain and the following lemma presents the conditional CDF of $R_t$ when $R_{t-1}$ is given.
\begin{lemma}
\label{SRLemmaConditionalPDF}
Given $R_{t} = u$, the conditional CDF of $R_{t+1}$, denoted by $\mathcal{P}_{t+1|t}^{\mathrm{SR}}\left(x|u\right)$, is
\begin{align}
\mathcal{P}_{t+1|t}^{\mathrm{SR}}\left(x|u\right)
 =\left\{
\begin{aligned}
\mathcal{P}_{0}^{\mathrm{SR}}\left(x|u\right) &\triangleq \left(1 - \left(\tilde{u}^{-1} x\right)^{-\frac{1}{q}}\right)\\
&\quad \times\mathbb{I}\left\{x \geq \tilde{u} \right\}, \quad\text{ if }\   t\geq\nu,\\
\mathcal{P}_{\infty}^{\mathrm{SR}}\left(x|u\right) &\triangleq \left(1 - \left(\tilde{u}^{-1} x\right)^{-\frac{1+q}{q}}\right)\\
&\quad \times \mathbb{I}\left\{x \geq \tilde{u}  \right\}, \quad\text{ if }\  t<\nu,
\end{aligned}
\right.
 \label{SRtestConditionalPDF}
\end{align}
where $\tilde{u} = \frac{1 + u}{1 + q}$.
\end{lemma}
\begin{IEEEproof}
Due to the fact that $\left|y_{t+1}\right|^2\geq0$, it is straight that $R_{t+1} \geq \frac{1 + u}{1 + q} $. We further have
\begin{align}
&\quad \mathcal{P}_{t+1|t}^{\mathrm{SR}}\left(x|u\right)=\mathbb{P}\left\{R_{t+1} \leq x | R_{t}=u\right\}
\nonumber \\
& =\mathbb{P}\left\{ \tilde{u} \mathrm{e}^{\frac{q}{1+q}\left|y_{t+1}\right|^2} \leq x \big| R_{t}=u\right\}\nonumber \\
& =\mathbb{P}\left\{\left|y_{t+1}\right|^2 \leq
\frac{1+q}{q}\ln\left( \tilde{u}^{-1} x\right) \bigg| R_{t}=u\right\}.
\end{align}
When $t\geq\nu$,  $\left|y_{t+1}\right|^2\sim\mathcal{E}\left(1+q\right)$ and thus $\mathbb{P}\left\{\left|y_{t+1}\right|^2 \leq x\right\} = 1 - \mathrm{e}^{-\frac{x}{1+q}}$, and when $t<\nu$, $\left|y_{t+1}\right|^2\sim\mathcal{E}\left(1\right)$ and $\mathbb{P}\left\{\left|y_{t+1}\right|^2 \leq x\right\} = 1 - \mathrm{e}^{-x}$, which leads to \eqref{SRtestConditionalPDF}.
\end{IEEEproof}

Next, we present a method to calculate $\mathcal{Q}_{L}^{\mathrm{SR}}(q)$. For $n \geq 1$ and $0\leq x < \eta_r$, we define
\begin{align}
\mathcal{Q}_n^{\mathrm{SR}}(x;q) \triangleq &\mathbb{P}\Big\{R_{\nu+1} < \eta_r,R_{\nu+2} < \eta_r,\nonumber \\
&\quad\quad \cdots,R_{\nu+n} < \eta_r | R_{\nu} = x , T_r > \nu\Big\}. \label{ConditionalSRQL}
\end{align}
By letting $n=L$ in \eqref{ConditionalSRQL}, we obtain the conditional covert probability $\mathcal{Q}_L^{\mathrm{SR}}(x;q)$ conditioned on $R_{\nu}=x$.
Therefore, $ \mathcal{Q}_{L}^{\mathrm{SR}}(q) = \mathbb{E}_{R_{\nu}}\left(\mathcal{Q}_L^{\mathrm{SR}}(R_{\nu};q)\right)$. To calculate $\mathcal{Q}_{L}^{\mathrm{SR}}(q)$, we first investigate $\mathcal{Q}_L^{\mathrm{SR}}(R_{\nu};q)$, and then focus on the probability distribution of $R_{\nu}$.

For $n=1$, we have $\mathcal{Q}_{1}^{\mathrm{SR}}(x;q)=1 - \left(\frac{1 + x}{1 + q}\right)^{\frac{1}{q}} \eta_r^{-\frac{1}{q}}$, which directly follows  from Lemma \ref{SRLemmaConditionalPDF}. When $n \geq 2$, $\mathcal{Q}_n^{\mathrm{SR}}(x;q)$ satisfies
\begin{align}
\label{RecursiveSRConditional}
&\quad \mathcal{Q}_n^{\mathrm{SR}}(x;q) \nonumber \\
&= \mathbb{P}\left\{R_{\nu+1} < \eta_r,\cdots,R_{\nu+n} < \eta_r
| R_{\nu} = x , T_r > \nu \right\} \nonumber \\
&=\int_{\frac{1+x}{1+q}}^{\eta_r} \mathbb{P}\Big\{R_{\nu+2} < \eta_r,R_{\nu+3} < \eta_r,\nonumber \\
&\quad\cdots,R_{\nu+n} < \eta_r
| R_{\nu+1} = y , R_{\nu} = x , T_r > \nu \Big\} \mathrm{d} \mathcal{P}_{0}^{\mathrm{SR}}\left(y|x\right)  \nonumber \\
&=\int_{\frac{1+x}{1+q}}^{\eta_r} \mathcal{Q}_{n-1}^{\mathrm{SR}}(y;q) \mathrm{d} \mathcal{P}_{0}^{\mathrm{SR}}\left(y|x\right),
\end{align}
where $\mathcal{P}_{0}^{\mathrm{SR}}\left(y|x\right)$ is defined in Lemma \ref{SRLemmaConditionalPDF}.
In fact, \eqref{RecursiveSRConditional} provides a recursive integral formula for $\mathcal{Q}_n^{\mathrm{SR}}(x;q)$.
Unfortunately, we are unable to obtain the closed-form expressions of $\mathcal{Q}_n^{\mathrm{SR}}(x;q)$ for $n\geq 2$ due to the complicated form of \eqref{RecursiveSRConditional}.
Nevertheless, we can obtain $\mathcal{Q}_n^{\mathrm{SR}}(x;q)$ numerically by using trapezoidal quadrature rule, see e.g. \cite{G.V.Moustakides2009,G.V.Moustakides2011,A.S.Polunchenko2014}.
To make the paper self-contained, we introduce a method to numerically calculate $\mathcal{Q}_n^{\mathrm{SR}}(x;q)$ in appendix \ref{Sec:QnxqNumCal}.

Next, we discuss how to compute the probability distribution of $R_\nu$. Under the condition that $T_r > \nu$, the distribution of $R_{\nu}$ is
\begin{align}
&\mathcal{G}_{\nu}^{\mathrm{SR}}(x;q) \triangleq \mathbb{P}\left\{ R_{\nu} \leq x | T_r > \nu \right\} \nonumber \\
=&
\mathbb{P}\left\{ R_{\nu}\leq x | R_{\nu} < \eta_r , R_{\nu-1} < \eta_r , \cdots, R_{1} < \eta_r \right\}.
\end{align}
For $1\leq n \leq \nu$, $\mathcal{G}_{n}^{\mathrm{SR}}(x;q)$ obeys the following recursive formula,
\begin{subequations}
\label{EqnGnuSR}
\begin{align}
\tilde{\mathcal{G}}_{n}^{\mathrm{SR}}(x;q)
&\triangleq
\mathbb{P}\left\{ R_{n}\leq x | T_r > n-1 \right\}
\nonumber \\
&= \int_{0}^{\eta_r} \mathcal{P}_{\infty}^{\mathrm{SR}}\left(x|y\right)  \mathrm{d} \mathcal{G}_{n-1}^{\mathrm{SR}}(y;q)\\
\mathcal{G}_{n}^{\mathrm{SR}}(x;q) &= \tilde{\mathcal{G}}_{n}^{\mathrm{SR}}(x;q)/\tilde{\mathcal{G}}_{n}^{\mathrm{SR}}(\eta_r;q),\nonumber \\
 &\quad \text{for } x \in \left(q^{-1}\left(1 - (1+q)^{-n}\right), \eta_r\right),
\end{align}
\end{subequations}
with $\mathcal{G}_{1}^{\mathrm{SR}}(x;q) =
\frac{1 - \left((1+q)x\right)^{- 1 - \frac{1}{q} }}{ 1 - \left((1+q)\eta_r \right)^{- 1 - \frac{1}{q} }}$ for $ \frac{1}{1+q} \leq x < \eta_r$.
We are unable to obtain a closed-form expression for $\mathcal{G}_{n}^{\mathrm{SR}}(x;q)$ for $n\geq 2$.
Nevertheless, we can numerically calculate $\mathcal{G}_{n}^{\mathrm{SR}}(x;q)$ following the method introduced in Appendix \ref{Sec:QnxqNumCal}.

Once $\mathcal{Q}_L^{\mathrm{SR}}(x;q)$ and
$\mathcal{G}_{\nu}^{\mathrm{SR}}(x;q)$ are obtained, we can numerically calculate $\mathcal{Q}_L^{\mathrm{SR}}(q) = \int  \mathcal{Q}_L^{\mathrm{SR}}(x;q) \mathrm{d} \mathcal{G}_{\nu}^{\mathrm{SR}}(x;q)$.
It is worth noting here that $\mathcal{Q}_{L}^{\mathrm{SR}}(q)$ depends on the value of $\nu$.

\subsection{Detection thresholds in the CUSUM and SR tests}
\label{ThresholdinCUSUMandSR}
Note that detection thresholds $\eta_{c}$ and $\eta_{r}$ in the CUSUM and SR tests should satisfy the ARL2FA constraint, i.e., $t_a = \gamma$. In this subsection, we present the method to determine the ARL2FA for a given detection threshold, based on which we can find the optimal detection threshold.

Define $\phi^{\mathrm{CU}}(x)=\mathbb{E}_{\infty}(T_c|\hat{C}_0 = x)$ and $\phi^{\mathrm{SR}}(x)=\mathbb{E}_{\infty}\left(T_r|R_0 = x\right)$, i.e., $\phi^{\mathrm{CU}}(u)$ ($\phi^{\mathrm{SR}}(u)$) is the ARL2FA of the CUSUM (SR) test under the condition that $\hat{C}_0$ ($R_0$) is initialized by $\hat{C}_0 = x$ ($R_0 = x$) with $0\leq x < \hat{\eta}_c$ ($0\leq x < \eta_r$). Obviously, $\phi^{\mathrm{CU}}(0) = t_a\left(T_c\right)$ and $\phi^{\mathrm{SR}}(0) = t_a\left(T_r\right)$.
Therefore, if $\phi^{\mathrm{CU}}(x)$ ($\phi^{\mathrm{SR}}(x)$) is derived, we can obtain the ARL2FA of the CUSUM (SR) test for a given threshold.
The expressions for $\phi^{\mathrm{CU}}(x)$ and $\phi^{\mathrm{SR}}(x)$ are presented in the following two lemmas.
\begin{lemma}[Eqn. (2.5) in \cite{S.Vardeman1985}]
\label{Lemma:CUSUMARL2FA}
Denote $M\triangleq \left\lceil \frac{\hat{\eta}_c}{\omega} \right\rceil$. For $n \omega \leq x < (n+1)\omega \wedge \hat{\eta}_c$ with $ 0\leq n \leq M-1 $, $\phi^{\mathrm{CU}}(x)$ is given by
\begin{align}
\phi^{\mathrm{CU}}(x) &= \phi^{\mathrm{CU}}(0) + 1 + n \nonumber \\
&\quad + \sum_{m = 0}^n (-1)^m c_{n-m}\frac{(x - (m-1)\omega)^m}{m!\mathrm{e}^{ -(x - m\omega) }},
\label{ARL2FACUSUM}
\end{align}
where for $n \geq 1$, $c_n$ is defined in \eqref{CUSUM-Cn-Iteration}
\begin{figure*}[t]
\begin{align}
c_{n} = c_{n-1} + \mathrm{e}^{-n\omega}\left(-1 + (-1)^n\frac{\omega^n}{ n! } + \sum_{m=1}^{n-1}\frac{(-1)^m(c_{n-1-m} - c_{n-m})\left((n-m+1)\omega\right)^m}{ m! \mathrm{e}^{- (n-m)\omega}}\right)\label{CUSUM-Cn-Iteration}
\end{align}
\noindent\rule[0.25\baselineskip]{\textwidth}{1pt}
\vspace{-7mm}
\end{figure*}
with $c_0 = -1$, and $\phi^{\mathrm{CU}}(0)$ can be obtained by solving linear equation $\phi^{\mathrm{CU}}(0)
= 1 + \int_{0}^{\hat{\eta}_c} \phi^{\mathrm{CU}}(x) \mathrm{d} \mathcal{P}_{\infty}^{\mathrm{CU}} (x|0)$ with respect to $\phi^{\mathrm{CU}}(0)$.
\end{lemma}
\begin{IEEEproof}
Please refer to \cite{S.Vardeman1985}.
\end{IEEEproof}

\begin{lemma}[Theorem 1 in \cite{W.Du2015}]
\label{Lemma:SRARL2FA}
If $\eta_r \geq \frac{1}{q}$, then for $x\in[0,\eta_r)$, $\phi^{\mathrm{SR}}(x)=1 + (1+q)\left(\eta_r - \frac{1 + x}{1+q}\right)$.
\end{lemma}
\begin{IEEEproof}
Please refer to \cite{W.Du2015}.
\end{IEEEproof}

Based on Lemma \ref{Lemma:CUSUMARL2FA}, the optimal detection threshold in the CUSUM test, i.e., $\hat{\eta}_c$, can be numerically obtained by using the bisection method. For the SR test, based on Lemma \ref{Lemma:SRARL2FA}, the optimal detection threshold is given by $\eta_r = \frac{\gamma}{1+q}$.

In summary, in this section, we derived the mathematical expressions for the covert probabilities in the Shewhart, CUSUM, and SR tests, respectively. The methods to obtain the optimal thresholds in the CUSUM and SR tests were also introduced.
Using these results, in the next section, we consider to optimize the transmit power and transmission duration in order to maximize the throughput while maintaining a desired covert probability.

\section{LPD communication performance evaluation}
In this section, we first discuss the condition under which LPD communication is feasible. Here, LPD communication is said to be feasible if the covert constraint can be satisfied by some $(q,L)$ pair with $q>0$ and $L\geq 1$.
Then, we present our method to optimize LPD communication performance under the covert probability constraint.
\subsection{Feasibile LPD communication}
For convenience, we refer to the communication system as a \emph{$(q,L)$-system} when Alice's transmit power and transmission duration are $q$ ($q>0$) and $L$ ($L\geq 1$), respectively.
For a given lower bound on the covert probability in \eqref{CovertCapacity}, i.e., $\theta$, a $(q,L)$-system is said to be \emph{$\theta$-covert} if the $(q,L)$ pair meets the covert constraint, i.e., $\mathcal{Q}_{L}(q)\geq \theta$.
The following theorem presents a necessary condition for a $(q,L)$-system being $\theta$-covert.
\begin{theorem}
\label{FesibilityForALL}
If a $(q,L)$-system is $\theta$-covert, then
\begin{align}
1 - \theta \geq
\left\{\begin{aligned}
& \left( 1/\gamma \right)^{\frac{1}{1+q}},~ \text{under Shewhart test},  \\
& \frac{\mathbb{E}\left(\mathrm{e}^{\frac{\hat{C}_\nu}{1+q}} \big| T_c> \nu\right)}{\mathrm{e}^{\frac{\omega+\hat{\eta}_c}{1+q}}},~\text{under CUSUM test},\\
& \mathbb{E}\left(
\left(\frac{1+R_{\nu}}{(1+q)\eta_r}\right)^{\frac{1}{q}}
\bigg |T_r>\nu\right),~\text{under SR test},
\end{aligned}\right.
\end{align}
\end{theorem}
\begin{IEEEproof}
As $\mathcal{Q}_{L}(q)$ is decreasing in $L$, if a $(q,L)$-system is $\theta$-covert, so does a $(q,1)$-system, i.e, $\mathcal{Q}_{1}(q) \geq \theta$, which leads to Theorem \ref{FesibilityForALL}. Note that if $L=1$, then the necessary condition in this theorem is also a sufficient condition.
\end{IEEEproof}

In fact, Theorem \ref{FesibilityForALL} provides us some insights on the feasibility of LPD communication that are summarized in the following three corollaries.
\begin{corollary}
\label{Cor:She}
Under the Shewhart test, there exists a $\theta$-covert $(q,L)$-system for $q>0$ and $L\geq 1$ if and only if $1 - \frac{1}{\gamma} > \theta$.
\end{corollary}
\begin{IEEEproof}
Based on Theorem \ref{FesibilityForALL},  $q \leq \frac{\ln(1/\gamma)}{\ln(1-\theta)} - 1$. As $q > 0$, we have $\frac{\ln(1/\gamma)}{\ln(1-\theta)} - 1 > 0$, which leads to Corollary \ref{Cor:She}.
\end{IEEEproof}
\begin{corollary}
Under the CUSUM test, there exists a positive number $\varepsilon>0$ such that if $1-\theta \leq \varepsilon$, any $(q,L)$-system cannot be $\theta$-covert.
\end{corollary}
\begin{IEEEproof}
Denote $\mathcal{U}_c(q)\triangleq \mathrm{e}^{-\frac{\omega+\hat{\eta}_c}{1+q}}
~\mathbb{E}\left(\mathrm{e}^{\frac{\hat{C}_\nu}{1+q}} | T_c > \nu\right)$.
Note that $\omega = \frac{1+q}{q}\ln(1+q)$ and $\hat{\eta}_c$ is the scaled detection threshold defined in \eqref{ScaledCUSUMtestSta}.
For a given ARL2FA of Willie's detector, $\gamma$, as $q\rightarrow 0^+$,  $\omega\rightarrow 1$ and $\hat{\eta}_c$ is bounded above (otherwise the ARL2FA becomes infinite). As a result, $\mathcal{U}_c(q)$ is lower bounded away from $0$ as $q\rightarrow 0^+$, i.e.,
$\lim_{q\rightarrow 0^+}  \mathcal{U}_c(q) >  \varepsilon \triangleq \lim_{q\rightarrow 0^+} \mathrm{e}^{-\frac{\omega+\hat{\eta}_c}{1+q}} >0$. Therefore, if $1 - \theta < \varepsilon$, any $(q,L)$-system cannot be $\theta$-covert even if Alice's transmit power approaches zero.
\end{IEEEproof}
\begin{corollary}
Under the SR test, if $\nu < \gamma - 2 - \mu$ for some positive real number $\mu$, then for any pre-given $\theta$ ($0<\theta<1$), there exists a $\theta$-covert $(q,L)$-system.
\end{corollary}
\begin{IEEEproof}
According to \eqref{RecursiveSRStatistic}, $R_{\nu} \overset{q\rightarrow 0}{\longrightarrow} 1 + R_{\nu-1}$. Therefore, as $q\rightarrow 0$, $R_{\nu} \rightarrow \nu < \nu + \frac{1}{2}\mu$.
In order for the ARL2FA to be $\gamma$, the detection threshold $\eta_r$ must satisfy $\eta_r > \gamma - 1$ as $q\rightarrow 0$.
Denote $\mathcal{U}_r(q) \triangleq \frac{1}{((1+q)\eta_r)^{\frac{1}{q}}}\mathbb{E}\left( (1+R_{\nu})^{\frac{1}{q}} \big |T_r>\nu\right)$, then $0\leq \lim_{q \rightarrow 0} \mathcal{U}_r(q) \leq \lim_{q \rightarrow 0} \frac{1}{(1+q)^{\frac{1}{q}}} \left(\frac{1 + \nu + \frac{1}{2}\mu}{\gamma - 1}\right)^{\frac{1}{q}}$.
Since $\lim_{q \rightarrow 0} \frac{1}{(1+q)^{\frac{1}{q}}} = \mathrm{e}^{-1}$ and $\frac{1 + \nu + \frac{1}{2}\mu }{\gamma - 1} < \frac{\gamma - 1 - \frac{1}{2}\mu}{\gamma - 1} < 1$ under the condition that $\nu < \gamma - 2 - \mu$, we have $\lim_{q \rightarrow 0} \mathcal{U}_r(q) = 0 < 1 - \theta $ for any $\theta$ ($0<\theta<1$).
\end{IEEEproof}

\begin{remark}
According to Corollary 1-3, we conjecture that if $\nu$ is relatively small compared with $\gamma$ and if $\theta$ is sufficiently large, then LPD communication performance under the SR test outperforms that under the Shewhart and CUSUM tests. This is because LPD communication becomes infeasible under the Shewhart and CUSUM tests if $\theta$ is sufficiently large as indicated by Corollary 1 and 2. We will verify this observation via numeric evaluation in Section \ref{Sec:Num}.
\end{remark}

\begin{figure*}[t]
  \centering
  \begin{minipage}[t]{\linewidth}
  \centering
  \subfigure[{\small $\mathcal{Q}_L(q)$ versus $L$ with $q = 0.15$.}]{
  \label{SimulateFig:11} 
  \includegraphics[width=2.5 in ]{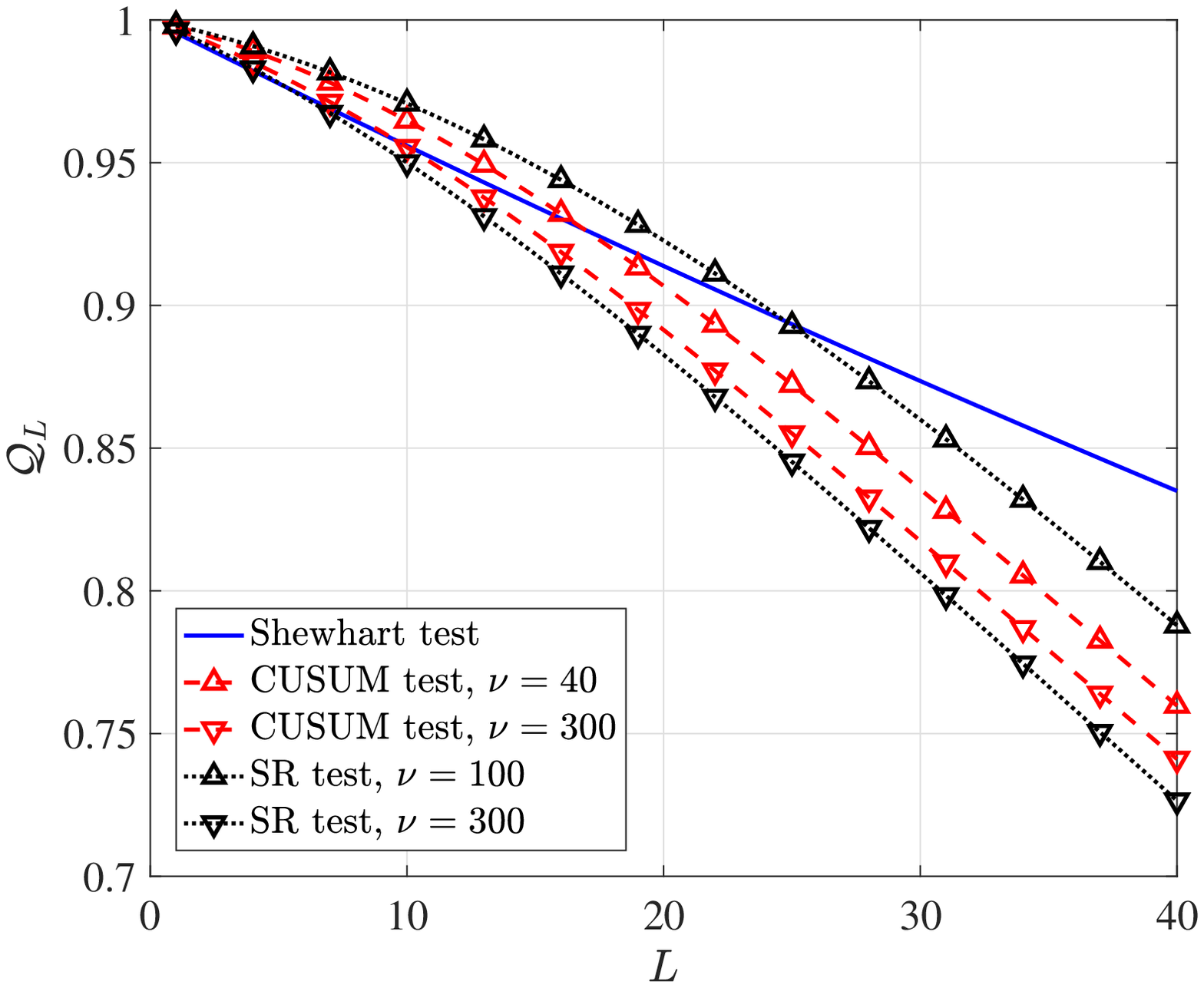}}
  \hspace{0.1 in}
  \subfigure[{\small $\mathcal{Q}_L(q)$ versus $q$ with $L = 15$.}]{
  \label{SimulateFig:12} 
  \includegraphics[width=2.5 in ]{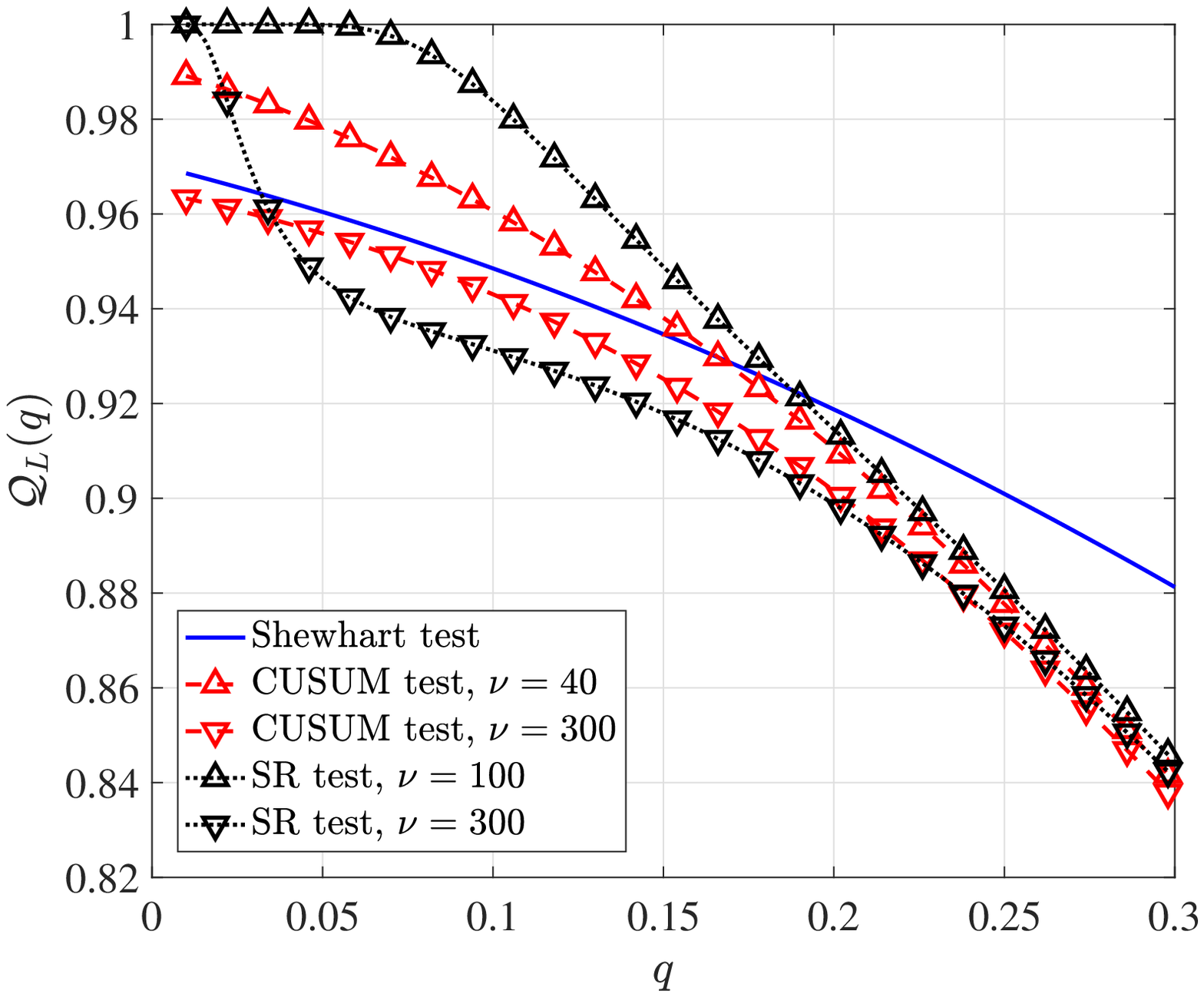}}
  \caption{\small Covert probability with $\gamma=500$.}
  \label{SimulateFig:1}
  \end{minipage}
  \vspace{-4mm}
\end{figure*}

\subsection{The Shewhart test}
\label{SHTestSolveProblem}
With $\mathcal{Q}_L^{\mathrm{SH}}(q)$ derived in Section \ref{QLDerivationSH}, we rewrite the maximization problem in \eqref{CovertCapacity} as follows,
\begin{align}
\begin{aligned}
\max_{L,q} &\quad L\ln\left(1 + q/\sigma_B^2\right),\\
\mathrm{s.t.}\ &\quad L \ln \left(1 - \gamma^{-\frac{1}{1+q}} \right)  \geq \ln \theta.\label{CovertCapacityInShewhartTest}
\end{aligned}
\end{align}
\begin{proposition}
\label{PropositionShewhartTest}
Denote $\left(q^*,L^*\right)$ as the optimal solution to \eqref{CovertCapacityInShewhartTest}.
The covert probability constraint in \eqref{CovertCapacityInShewhartTest} is active at the optimum, i.e., $ {L^*} \ln \left(1 - \gamma^{-\frac{1}{1+q^*}} \right) = \ln \theta$.
\end{proposition}
\begin{IEEEproof}
This is because the objective function in \eqref{CovertCapacityInShewhartTest} is  monotonically increasing in both $q$ and $L$, and $\mathcal{Q}_L^{\mathrm{SH}}(q)$ is monotonically decreasing in both $q$ and $L$.
\end{IEEEproof}

Based on Proposition \ref{PropositionShewhartTest}, for a fixed $L$, the optimal transmit power is
\begin{align}
q = - 1 +  \frac{\ln\left(1/\gamma\right) }{\ln\left(1 - \theta^{1/L}\right) } .\label{OptimalPinShewhartTest}
\end{align}
Therefore, the maximization problem in \eqref{CovertCapacityInShewhartTest} becomes
\begin{align}
\max_{L} & \quad L\ln\left(1 - \frac{1}{\sigma_B^2} + \frac{ \ln(1/\gamma)}{\sigma_B^2\ln\left(1 - \theta^{1/L}\right)}\right).\label{EquivalentL}
\end{align}
Since $q\geq0$, it follows from \eqref{OptimalPinShewhartTest} that
$L\leq L_{\max} \triangleq  \ln(\theta) /\ln\left(1 - \gamma^{-1}\right)$.
Consequently, the solution to \eqref{EquivalentL}, $L^*$, lies in $\{1,2,\cdots,\left\lfloor L_{\max} \right\rfloor \}$ and can be obtained by an exhaustive search, where $\lfloor x\rfloor$ is the floor function.
Note that $L_{\max}$ increases with $\gamma$, which means the search range for solving \eqref{EquivalentL} increases with $\gamma$. In the following, we propose a closed-form approximate solution for \eqref{CovertCapacityInShewhartTest}.

According to \eqref{OptimalPinShewhartTest}, for any fixed $q$, the optimal transmission duration is
$L^*\left(q\right) = \left\lfloor  \ln(\theta)/\ln\left(1 - \gamma^{-\frac{1}{1+q}}\right)   \right\rfloor$.
Substituting $L^*\left(q\right)$ to \eqref{CovertCapacityInShewhartTest}, problem \eqref{CovertCapacityInShewhartTest} becomes
\begin{align}
\max_{q} \quad
\left\lfloor  \ln(\theta)/\ln\left(1 - \gamma^{-\frac{1}{1+q}}\right)   \right\rfloor
\ln\left(1 + q/\sigma_B^2\right). \label{ObjectiveInShewhartTestOnlyP}
\end{align}
Neglecting the floor function in \eqref{ObjectiveInShewhartTestOnlyP}, we obtain $\check{\mathcal{I}}(q)\triangleq \frac{\ln(\theta)}{\ln\left(1 - \gamma^{-1/(1+q)}\right)}\ln\left(1 + q\frac{\sigma_W^2}{\sigma_B^2}\right)$.
Note that  $\gamma$ is usually chosen to be large enough to avoid frequent false alarms, and in covert transmission, transmit power is generally small,
therefore, $\check{\mathcal{I}}(q)\approx\hat{\mathcal{I}}(q)\triangleq -\frac{\ln(\theta)\sigma_W^2}{\sigma_B^2}q\gamma^{\frac{1}{1+q}}$.

By checking the first order derivative, we find that $\hat{\mathcal{I}}(x)$ first reaches a local maximum, at which $x = u_1 \triangleq \frac{-(2 - \ln(\gamma)) -\sqrt{(2 - \ln(\gamma))^2 - 4}}{2}$, then decreases with respect to $x$ until it reaches a local minimum,  at which  $x = u_2 \triangleq \frac{-(2 - \ln(\gamma)) +\sqrt{(2 - \ln(\gamma))^2 - 4}}{2}$. In the region of $x>u_2$, $\hat{\mathcal{I}}(x)$ monotonically increases and $\hat{\mathcal{I}}(x)\rightarrow+\infty$ as $x\rightarrow+\infty$.
However, different from $\hat{\mathcal{I}}\left(x\right)$, $\mathcal{I}(x,L(x))\rightarrow 0 $ as $x\rightarrow+\infty$. Therefore, we propose to use $\tilde{q}^* = u_1$ as a rough approximation for the optimal transmit power. In this way, the optimal $L$ to the  problem in \eqref{CovertCapacityInShewhartTest} can be approximated by $\hat{L}^* = L\left(\tilde{q}^*\right)$.
With $\hat{L}^*$ at hand, we can obtain a refined approximation of the optimal transmit power $\hat{q}^* = \frac{-\ln(\gamma)}{\ln\left(1 - \theta^{1/\hat{L}^*}\right)} - 1$, which follows from \eqref{OptimalPinShewhartTest}.  The accuracy of the approximate solution will be shown in Section \ref{Sec:Num}.

\subsection{The CUSUM and SR tests}

\begin{figure*}[t]
  \centering
  \includegraphics[width=6.5 in]{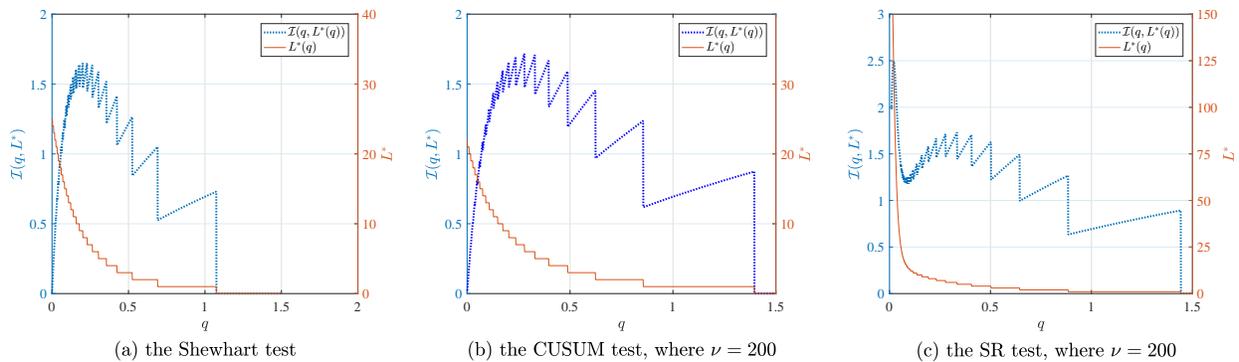}\\
  \caption{\small $\mathcal{I}(q, L^*(q))$ and $L^*(q)$ versus $q$.}\label{FIG:IVSq}
  \vspace{-2mm}
\end{figure*}
\begin{figure*}[t]
  \centering
  \includegraphics[width=6.5 in]{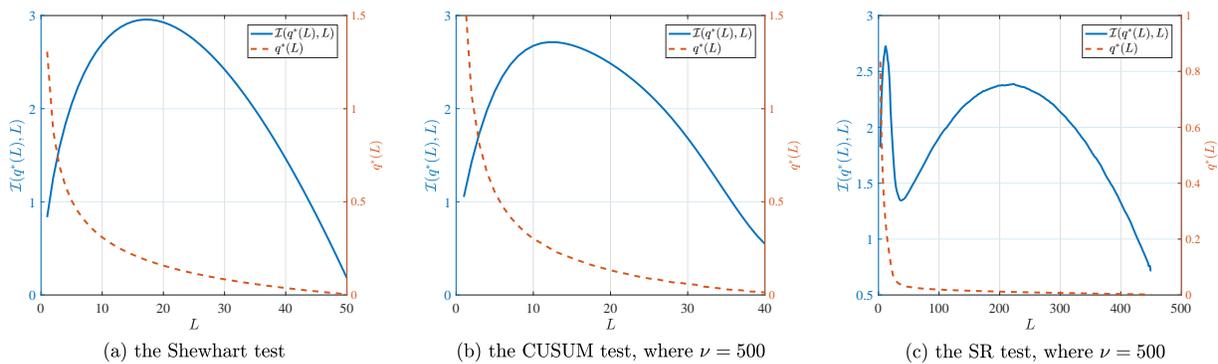}\\
  \caption{\small $\mathcal{I}(q^*,L)$ versus $L$ with $\theta = 0.95$ and $\gamma = 1000$.}\label{Fig:IL}
  \vspace{-2mm}
\end{figure*}

In this subsection, we present our method to evaluate LPD communication performance under the condition that Willie performs the CUSUM or SR test.
Due to the complicated mathematical formulas for  the covert probabilities in CUSUM and SR tests, it is hard to obtain analytical solution to problem \eqref{CovertCapacity}.
Therefore, in the following, we solve \eqref{CovertCapacity} numerically.

In fact, the major difficulty in solving \eqref{CovertCapacity} lies in how to deal with the covert probability constraint, $\mathcal{Q}_L^{\mathrm{X}}(q) \geq  \theta$, where we denote $\mathrm{X} \in \{\mathrm{CU},\mathrm{SR}\}$ for notational simplicity.
Since we only have two optimization variables, i.e., $q$ and $L$, we can first fix one of them and  optimize the other. In this way, \eqref{CovertCapacity} degrades to an optimization problem with a single optimization variable, and  can be solved by an one-dimensional search.
Regarding $\mathcal{Q}_L^{\mathrm{X}}(q)$, we observe that $\mathcal{Q}_L^{\mathrm{X}}(q)$ decreases with $L$ if $q$ is fixed.
Besides, when $q$ is fixed,  the calculation of $\mathcal{Q}_l^{\mathrm{X}}(q)$ with $l>1$ can be based on  intermediate results obtained during the calculation of $\mathcal{Q}_{l-1}^{\mathrm{X}}(q)$, i.e., $\mathcal{Q}_{l-1}^{\mathrm{X}}(x;q)$. This means that it is computationally much easier to find the optimal $L$ for fixed $q$ than to find the optimal $q$ for fixed $L$.
Therefore, we propose to optimize $L$ first with $q$ fixed as a constant, and then optimize $q$ by an exhaustive one-dimensional search. We summarize our method to solve \eqref{CovertCapacity} when Willie performs the CUSUM or the SR test in Algorithm \ref{TABLE:DistanceBasedMethod}, wherein $(q_{\min},q_{\max})$ is the searching interval and $\Delta q$ is the step size.

\begin{remark}
Note that in this paper, the covertness of the communication between Alice and Bob is evaluated through covert probability.
If we let  Willie optimize his detection strategy based on the transmit power and transmission duration of Alice to minimize the covert probability, then Alice's problem becomes
\begin{subequations}
\label{GameTCovertCapacity}
\begin{align}
\max_{q,L}& \quad \mathcal{I}\left(q,L\right),\label{GameTCovertCapacity-a}\\
\mathrm{s.t.}& \quad \mathcal{Q}_{L}^{(\nu)}(q;T(q,L)) \geq \theta,
\label{GameTCovertCapacity-b} \\
&\quad
T(q,L) = \mathop{\mathrm{argmin}}\limits_{T;\mathbb{E}_{\infty}(T)\geq \gamma} \sup_{\nu} \mathcal{Q}_{L}^{(\nu)}(q;T),\label{GameTCovertCapacity-c}
\end{align}
\end{subequations}
where
$T(q,L) = \inf\{t: t\geq 1, Z_t(q,L)\geq \eta_t(q,L)\}$ is the stopping time to be optimized by Willie with $Z_t(q,L)$ and $\eta_t(q,L)$ being the test statistic and the detection threshold at time $t$, respectively,
and $\mathcal{Q}_{L}^{(\nu)}(q;T)$ is the covert probability for a given group of $\{\nu,q,L,T\}$.
Note that in \eqref{GameTCovertCapacity}, $\nu$ is known to Alice and Bob but is unknown to Willie, and the supremum operation in \eqref{GameTCovertCapacity-c} follows from the widely used minimax principle in the design of non-Baysian SCPD algorithm due to the unknown change-point $\nu$, see e.g., \cite[Chapter 6]{A.G.Tartakovsky2013}.
The major difference between \eqref{CovertCapacity} and \eqref{GameTCovertCapacity} is that $T$ is restricted to be one of the Shewhart, the CUSUM, and the SR tests in \eqref{CovertCapacity}, while in \eqref{GameTCovertCapacity}, the covert probability is minimized with respect to $T$  for any given pair of $(q,L)$.
By solving \eqref{GameTCovertCapacity}, we  may achieve an equilibrium between the communication performance achieved by Alice and Bob and the detection performance of Willie.
However, although the optimization problem in \eqref{GameTCovertCapacity} is expressed in such a concise form, it is very hard to solve.
In fact, the optimization problem in \eqref{GameTCovertCapacity} consists of two layers.
In the outer-layer problem, $\mathcal{I}(q,L)$ is maximized with respect to $(q,L)$ subject to the covert probability constraint in \eqref{GameTCovertCapacity-b}.
In the inner-layer problem, the covert probability is minimized with respect to $T$, which includes the optimization of the structure of the detection statistic $Z_t$ and the value of the detection threshold $\eta_t$.
The first difficulty in solving \eqref{GameTCovertCapacity} lies in the fact that
the solution to the inner-layer problem in \eqref{GameTCovertCapacity-c} is generally unknown.
To the best knowledge of the authors, \eqref{GameTCovertCapacity-c} can be theoretically solved only in the case $L=1$, and the optimal solution is the well-known Shewhart test, see \cite{M.Pollak2013,G.V.Moustakides2014}. However, for $L\geq 2$, \eqref{GameTCovertCapacity-c} is still an open problem.
The second difficulty is that
even if \eqref{GameTCovertCapacity-c} can be numerically solved, we need to compute it for each different pair of $(q,L)$  when numerically searching the optimal $(q,L)$ pair, the computational complexity of which is high.
Based on this two reasons, we are not able to solve \eqref{GameTCovertCapacity} currently, but solving this problem  constitutes an interesting  future research issue. In this paper, we assume that Willie adopts one of the three sequential tests, i.e., the Shewhart, the CUSUM, and the SR tests.
As a result, problem \eqref{GameTCovertCapacity} degrades to \eqref{CovertCapacity}, which is mathematically tractable.
Nevertheless, the results obtained in this paper are still useful as they can be viewed as theoretical upper bounds on the achievable LPD communication performance under sequential detection, which has not been particularly explored in the literature.
\end{remark}
\begin{algorithm}[t]
\caption{\small Numeric method to solve \eqref{CovertCapacity} when Willie performs the CUSUM (or SR) test.}
\label{TABLE:DistanceBasedMethod}
\begin{algorithmic}[1] 
\State \textbf{\small Input}:\  $\nu$, $\theta$, $q_{\max}$, $q_{\min}$, $\Delta q$
\State \textbf{\small Initialize}:\  $\mathcal{I}_{\mathrm{opt}}=0$, $q_{\mathrm{opt}} = 0$, $L_{\mathrm{opt}} = 0$;
\State $q = q_{\min}$;
\State \textbf{\small Repeat}:
\State \quad $l = 1$;
\State \quad Calculate $\mathcal{G}_{\nu}^{\mathrm{X}}(x;q)$ according to Theorem \ref{Theorem:StationaryCUSUMdistribution} (or \eqref{EqnGnuSR});
\State \quad \textbf{\small Repeat}:
\State \quad\quad Calculate $\mathcal{Q}_l^{\mathrm{X}}(x;q)$ according to Theorem \ref{Therorem:CUSUMQL} (or \eqref{RecursiveSRConditional});
\State \quad\quad Calculate $\mathcal{Q}_l^{\mathrm{X}}(q)$ using $\mathcal{Q}_l^{\mathrm{X}}(x;q)$ and $\mathcal{G}_{\nu}^{\mathrm{X}}(x;q)$;
\State \quad\quad $\mathcal{I} = l \times \ln (1 + (\sigma_W^2/\sigma_B^2)q)$;
\State \quad\quad \textbf{\small If} $\mathcal{I}>\mathcal{I}_{\mathrm{opt}}$ \& $\mathcal{Q}_l^{\mathrm{X}}(q) \geq \theta$ \textbf{\small Then} $\mathcal{I}_{\mathrm{opt}}=\mathcal{I}$, $q_{\mathrm{opt}} = q$, $L_{\mathrm{opt}} = l$;
\State \quad\quad $l = l + 1$;
\State \quad \textbf{\small Until} $\mathcal{Q}_l^{\mathrm{X}}(q) < \theta$;
\State \quad $q = q + \Delta q$;
\State \textbf{\small Until} $q > q_{\max}$;
\end{algorithmic}
\end{algorithm}

\section{Numeric Results and Discussions}
\label{Sec:Num}
\begin{figure*}[t]
  \begin{center}
  \centering
  \includegraphics[width=6.5 in]{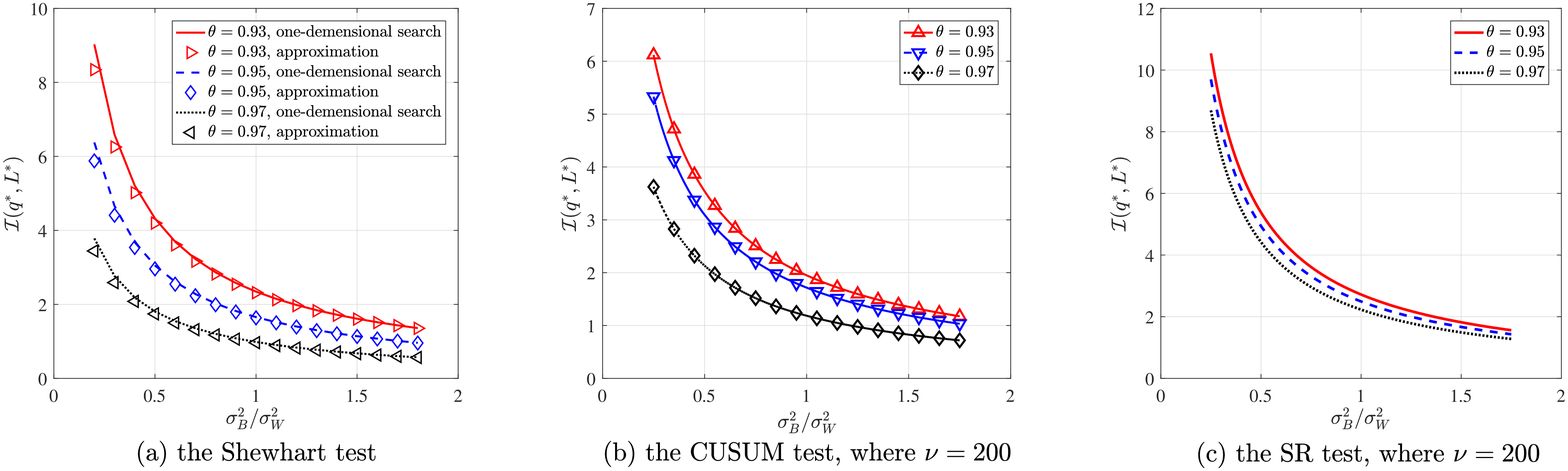}
  \caption{\small $\mathcal{I}(q^*,L^*)$ versus the ratio between Bob's and Willie's noise power.}\label{INOISE}
  \end{center}
  \vspace{-2mm}
  \begin{minipage}[t]{0.48\linewidth}
  \centering
  \includegraphics[width=2.8 in]{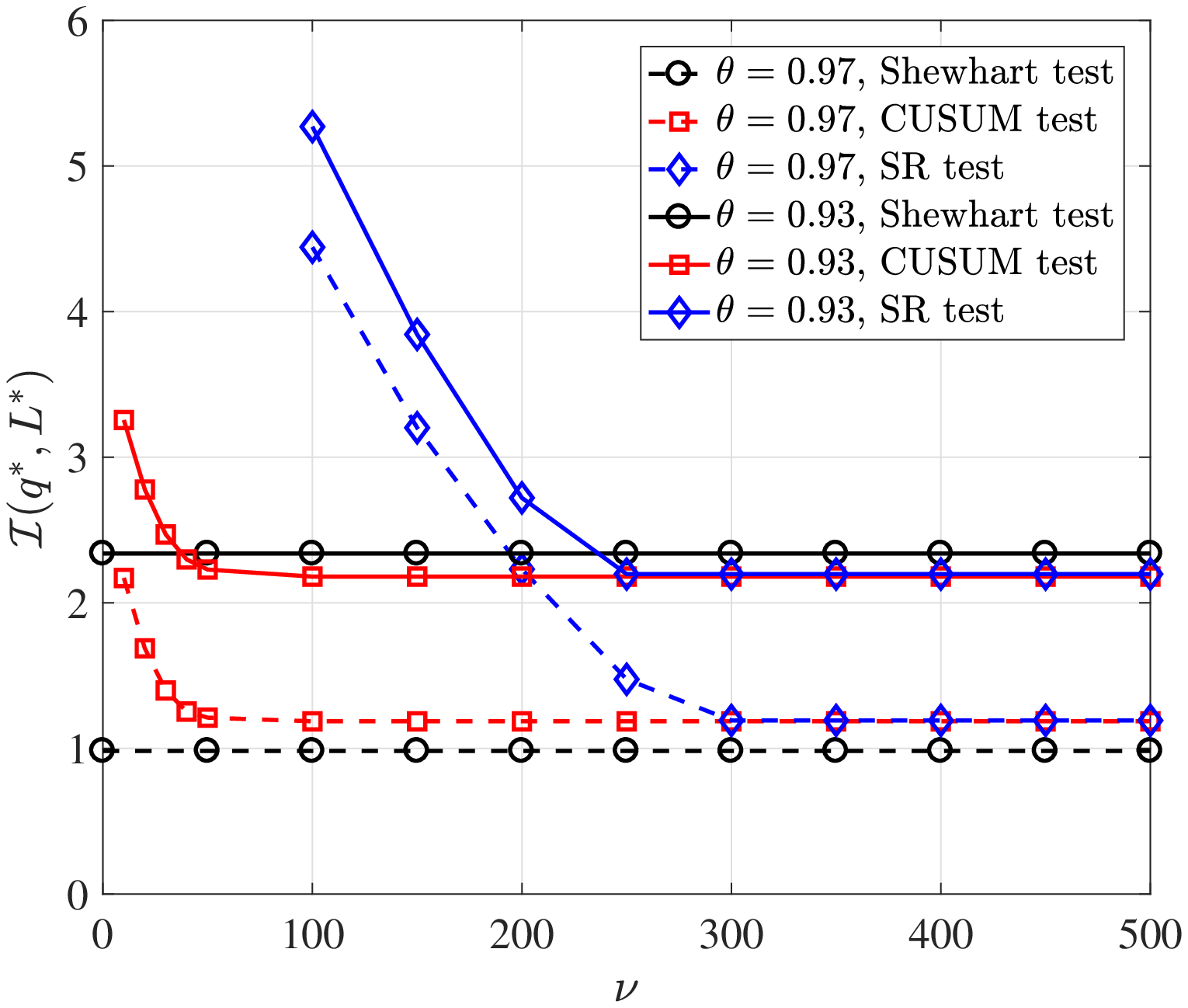}
  \caption{\small $\mathcal{I}({q^*,L^*})$ versus $\nu$}\label{SimulateFig:41}
  \end{minipage}
  \begin{minipage}[t]{0.48\linewidth}
  \centering
  \includegraphics[width=2.8 in]{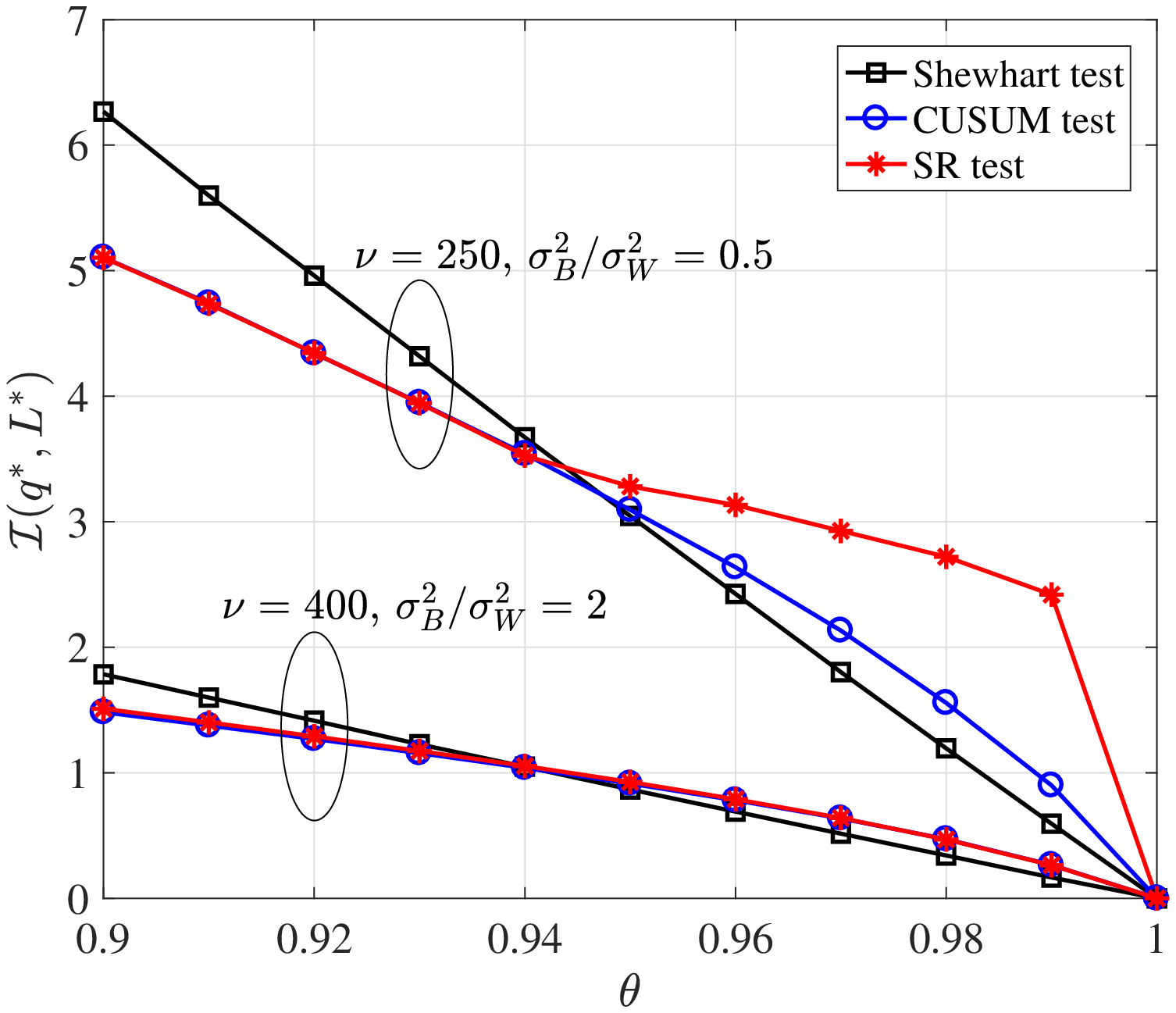}
  \caption{\small $\mathcal{I}({q^*,L^*})$ versus $\theta$.}\label{SimulateFig:42}
  \end{minipage}%
  \vspace{-2mm}
\end{figure*}

\subsection{Numeric results}
In this part, we numerically evaluate the performance of the LPD communication. Unless specified, in the simulation, we set the ARL2FA of Willie's detector $\gamma = 500$, the lower bound on the covert probability $\theta = 0.95$, and $\sigma_B^2/\sigma_W^2 = 1$.

In Fig. \ref{SimulateFig:11} and Fig. \ref{SimulateFig:12}, we plot the covert probabilities versus the transmission duration and the transmit power, respectively, under the Shewhart, CUSUM, and SR tests.
Fig. \ref{SimulateFig:1} reveals that increasing either the transmit power or the  transmission duration decreases the covert probability. Therefore, to transmit as much information content as possible subject to a sufficiently high covert probability, it is necessary to carefully select the transmit power and transmission duration.
In Fig. \ref{SimulateFig:1}, we also show that  covert probabilities under CUSUM and SR tests are affected by $\nu$, i.e., the time when LPD communication occurs,
and a larger value of $\nu$ results in a smaller covert probability.
This is because the detector accumulates more noise samples before Alice starts to transmit to Bob as $\nu$ increases.
Unlike the CUSUM and SR tests, the covert probability in the Shewhart test does not depend on $\nu$ due to the fact that the Shewhart test statistic only depends on the instantaneous signal received at each moment.

We plot the optimal transmission duration for fixed $q$, denoted by $L^*(q)$, versus $q$ under  the Shewhart, CUSUM, and   SR tests in Fig. \ref{FIG:IVSq}(a), \ref{FIG:IVSq}(b), and  \ref{FIG:IVSq}(c), respectively.
From Fig. \ref{FIG:IVSq}, we can see that $L^*(q)$ is  non-increasing and eventually converges to $0$.
Based on $L^*(q)$, $\mathcal{I}(q,L^*(q))$
is illustrated as a function of $q$ in Fig. \ref{FIG:IVSq}.
In our system settings, under the condition that Willie performs the  Shewhart or CUSUM test, the envelope of $\mathcal{I}(q,L^*(q))$  first increases  and then decreases, see Fig. \ref{FIG:IVSq}(a) and \ref{FIG:IVSq}(b) respectively.
This increase-first-and-then-decrease property indicates that we need to search for the optimal transmit power that maximizes $\mathcal{I}(q,L^*(q))$.
The curve of $\mathcal{I}(q,L^*(q))$ under the SR test appears to be slightly different. As shown in Fig. \ref{FIG:IVSq}(c), the envelope of $\mathcal{I}(q,L^*(q))$ has two local maxima.
One of them suffers a low transmit power level but has a long period of time that can be utilized for covert transmission.
The other one has a sightly higher transmit power level and thus the feasible transmission duration is shorter.

In Fig. \ref{Fig:IL},
we plot $\mathcal{I}(q^*(L),L)$ versus $L$, where $q^*(L)$ denotes the optimal transmit power for fixed $L$. We assume that Willie performs the Shewhart, the CUSUM, and the SR tests in Fig. \ref{Fig:IL}(a), \ref{Fig:IL}(b), and \ref{Fig:IL}(c), respectively. From Fig. \ref{Fig:IL}, we can see that in all cases, $\mathcal{I}(q^*(L),L)$ tends to be decreasing with respect to $L$ as long as $L$ is sufficiently large. This means that simply increasing the transmission duration does not always improve the performance of LPD communication. We  point out that this conclusion differs from that in \cite{S.YanITFS2019}. In \cite{S.YanITFS2019}, it was shown that if the transmission time was longer, then LPD communication performance became better, but in out case, $\mathcal{I}(q^*(L),L)$ is maximized in the middle of the feasible region of $L$.

\begin{figure*}[t]
  \centering
  \includegraphics[width=6.5 in]{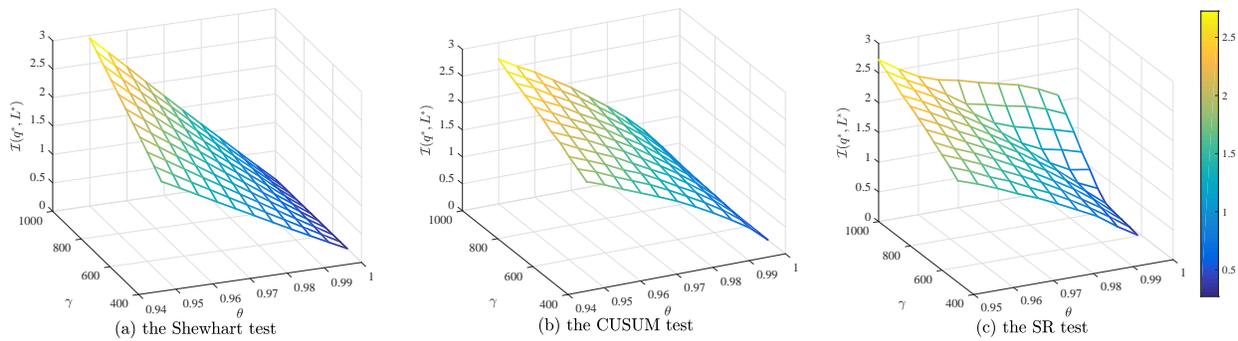}\\
  \caption{\small $\mathcal{I}(q^*,L^*)$ versus $\gamma$ with $\theta\in [0.95,0.99]$ and $\nu = 500$.}\label{SimulateFig:5}
  \vspace{2mm}
\end{figure*}

In Fig. \ref{INOISE}(a), \ref{INOISE}(b), and \ref{INOISE}(c), we illustrate $\mathcal{I}(q^*,L^*)$ versus the ratio of Bob's and Willie's noise powers under the Shewhart, CUSUM, and SR tests, respectively.
In Fig. \ref{INOISE}(a), the approximation derived in Section \ref{SHTestSolveProblem} is compared to results obtained from an exhaustive one-dimensional search. From Fig. \ref{INOISE}(a), we conclude that the approximation approaches to the optimal performance.
The curves in Fig. \ref{INOISE}(b) and \ref{INOISE}(c) are obtained numerically following Algorithm \ref{TABLE:DistanceBasedMethod}, where we set $q_{\max} = 2$, and $q_{\min} = \Delta q = 10^{-3}$.
In fact, Fig. \ref{INOISE} reveals two fundamental methods to enhance LPD communication performance, i.e., to suppress the noise of Bob's receiver (or equivalently to enhance the received signal power) and to inject random noise into Willie's receiver.

As the covert probabilities $\mathcal{Q}_{L}^{\mathrm{CU}}(q)$ and $\mathcal{Q}_{L}^{\mathrm{SR}}(q)$ are influenced by $\nu$ (see Fig. \ref{SimulateFig:1}), we evaluate LPD communication performance for different values of $\nu$, which are plotted in Fig. \ref{SimulateFig:41}.
Note that $\mathcal{I}({q^*,L^*})$ under the Shewhart test is also plotted, and as $\mathcal{Q}_{L}^{\mathrm{SH}}(q)$ does not rely on $\nu$, the curves for the Shewhart test are horizontal lines.
From Fig. \ref{SimulateFig:41}, it is observed that under the CUSUM and  SR tests, $\mathcal{I}({q^*,L^*})$ monotonically decreases with $\nu$ until it converges.
In fact, $\mathcal{Q}_{L}^{\mathrm{CU}}(q)$ and $\mathcal{Q}_{L}^{\mathrm{SR}}(q)$ decrease with $\nu$, which can be observed from Fig. \ref{SimulateFig:1}. This phenomenon leads to the fact that an increase in $\nu$ leads to a decrease in the maximum feasible transmission time length, and therefore, $\mathcal{I}({q^*,L^*})$ decreases.
Besides, as $\nu$ increases, the conditional distribution of Willie's test statistic at time $\nu$, i.e., $\mathcal{G}_{\nu}^{\mathrm{X}}(x;q)$ for $\mathrm{X}\in\{\mathrm{SR},\mathrm{CU}\}$,  converges to a quasi-stationary distribution, see \cite{A.G.Tartakovsky2013}.
Once $\mathcal{G}_{\nu}^{\mathrm{X}}(x;q)$ has converged, the covert probability will not change with $\nu$ anymore, and $\mathcal{I}({q^*,L^*})$ converges to a constant.
Hence, Fig. \ref{SimulateFig:41} suggests that having a sufficiently large value of $\nu$ produces a lower bound on the actual performance.
From Fig. \ref{SimulateFig:41}, we also observe that for small $\nu$, $\mathcal{I}({q^*,L^*})$ under the CUSUM test is smaller than that under the SR test. This because SR test is not sensitive to a statistical change that occurs at the beginning of the detection procedure, and for small $\nu$, the SR test usually allows a long period time for LPD communication.
As long as $\nu$ is sufficiently large, $\mathcal{I}({q^*,L^*})$ becomes similar under the CUSUM and SR tests.

In Fig. \ref{SimulateFig:42}, we illustrate $\mathcal{I}({q^*,L^*})$ versus the covert probability constraint $\theta$.
In fact, the curves in Fig. \ref{SimulateFig:42} can be viewed as a trade-off between the throughput and risk of being detected, and with the increase of $\theta$,  $\mathcal{I}({q^*,L^*})$ finally decreases to $0$.
Fig. \ref{SimulateFig:42} also reveals the distinct communication performance that can be achieved under the three sequential tests.
From Fig. \ref{SimulateFig:42}, we observe that if we desire a low risk of being detected, i.e., $\theta$ is close to $1$, then the Shewhart test restricts Alice to the lowest throughput while the SR test allows the largest value of $\mathcal{I}({q^*,L^*})$.
However, the contrary is the case when the covert probability constraint is loose.
As shown in Fig. \ref{SimulateFig:42}, when $\theta$ is around $0.92$, $\mathcal{I}({q^*,L^*})$ is smaller under CUSUM and SR tests than under Shewhart test.

In Fig. \ref{SimulateFig:5}, we plot $\mathcal{I}({q^*,L^*})$ versus $\gamma$ with $\theta\in[0.95,0.99]$.
Fig. \ref{SimulateFig:5} reveals that $\mathcal{I}({q^*,L^*})$ is monotonically increasing with $\gamma$.  Recall that $\gamma$ is the ARL2FA of Willie's detector. As we have mentioned in Section II, the value of $\gamma$ depends on the ability of Willie to tolerate the false alarms. If Willie can tolerate frequent false alarms, then he has a small the value of $\gamma$. But if Willie desires few false alarm, then the value of $\gamma$ must be large, which inevitably results in long detection delay. Intuitively, an increase of the detection delay allows Alice to transmit for a longer period of time, and therefore transmit more information covertly.

\subsection{Discussions}
\label{DifferenceFinal}
We note that the LPD communication scheme investigated in this paper differs from schemes studied in existing works.
Specifically, the differences mainly lie in the following three aspects:
\begin{enumerate}
\item \emph{Detection process:} in earlier works, the detection process of Willie was formulated as a BHT problem, and a final decision was made after Alice's transmission session terminated. In this paper, Willie makes sequential decisions by running an SCPD algorithm  due to the lack of the knowledge about when Alice's transmission starts;

\item \emph{Willie's goal:} in earlier works, the goal of Willie can be understood as determining whether or not communication occurred during a given time window that Willie is interested in; while in this paper,  Willie aims at discovering Alice's transmission as soon as possible once it starts;

\item \emph{Definition of covertness:} in earlier works, communication between Alice and Bob was said to be covert if the sum of the false alarm and missed detection probabilities approaches one. Such a metric for covertness cannot be directly extended to our case. In this paper, by excluding the impact of the false alarms (as explained in Section \ref{Section:LPDProState}), Alice's transmission is regarded to be covert if the event of missed detection, as defined in Fig. \ref{Model3Cases}, occurs with a sufficiently high probability, or equivalently, if the probability of the timely detection event is small.
\end{enumerate}
Because of these differences, the application scenarios of the LPD communication schemes studied in earlier works and in this paper are different. Specifically, the scheme studied in earlier works is useful in protecting privacy, for example, preventing Willie from analyzing the traffic pattern between Alice and Bob,
while the scheme presented in this paper can be useful in the situations where Willie wants to perform active attack toward ongoing transmission from Alice to Bob.

\section{Conclusions and Future Directions}
\label{Sec:FinalSection}

In this paper, we have established a new framework for LPD communication by formulating Willie's detection problem as an SCPD problem.
Three different SCPD algorithms, i.e., the Shewhart, the CUSUM, and the SR tests, have been considered to enable Willie's detection. For each of them, we have analyzed the covert probability, based on which we have further maximized the throughput subject to a sufficiently high covert probability by optimizing the transmit power and transmission duration.
Numerical results have been presented to show LPD communication performance, which reveal that:
1) it is viable to achieve LPD communication by utilizing the detection delay of Willie, and we can carefully select the transmit power and the transmission duration to achieve the optimal performance;
2) the ratio of Bob's and Willie's noise powers affects LPD communication performance a lot, and two basic methods to improve LPD communication performance are to suppress the noise of Bob's receiver and to inject interference into Willie's receiver;
and
3) the performance of LPD communication is significantly influenced by the time difference between when Willie starts its detection procedure and when Alice starts to transmit if Willie performs the CUSUM or the SR tests.

The new LPD communication framework established in this paper opens up several future research directions.
For example, in this paper, Alice is assumed to know the ARL2FA of Willie's detector. In general, the ARL2FA of Willie's detector is used to control the costs due to the false alarms. If the problem of LPD communication can be modeled as a game between Alice and Willie, then it might be possible to drop such assumption, and the ARL2FA might be directly determined by the equilibrium of the game.
Besides, it is interesting to study the performance improvement of LPD communication when Willie's sequential detector suffers from the imperfect estimation on the noise power, e.g., by using the noise uncertainty model in \cite{S.LeeJSTSP2015,B.HeCL2017}.
Another possible research issue is to investigate how to fully deteriorate the detection performance of Willie's SCPD. Intuitively, we may elaborate the transmit power or design a transmission scheme to constrain that the difference between ARL2FA and average detection delay of Willie's detector is small. In this way, the Willie's detector becomes ineffective if Willie wants to timely detect the occurrence of the communication.

\appendix
\subsection{The derivation of Theorem \ref{Therorem:CUSUMQL}}
\label{Appendix:A}
According to \eqref{ConditionalCUSUMQL}, for $0\leq x < \hat{\eta}_c$, $\mathcal{Q}_n^{\mathrm{CU}}(x)$ satisfies the recursive formula in \eqref{DerivationofCUSUMConditionalQn},
\begin{figure*}[t]
\begin{align}
\mathcal{Q}_{n+1}^{\mathrm{CU}}(x;q)
&= \mathbb{P}\left\{\hat{C}_{\nu+1} < \hat{\eta}_c,\hat{C}_{\nu+1} < \hat{\eta}_c,\cdots,\hat{C}_{\nu+n+1} < \hat{\eta}_c | \hat{C}_{\nu} = x, T_c > \nu \right\} \nonumber\\
&\overset{(a)}{=} \int_{0}^{\hat{\eta}_c}\mathbb{P}\left\{\hat{C}_{\nu+2} \leq \hat{\eta}_c,\cdots,\hat{C}_{\nu+n+1} \leq \hat{\eta}_c | \hat{C}_{\nu+1} = y ,\hat{C}_{\nu} = x, T_c>\nu\right\}\mathrm{d}\mathbb{P}\left\{\hat{C}_{\nu+1} \leq y | \hat{C}_{\nu} = x\right\} \nonumber \\
&\overset{(b)}{=} \int_{0}^{\hat{\eta}_c}\mathbb{P}\left\{\hat{C}_{\nu+2} \leq \hat{\eta}_c,\cdots,\hat{C}_{\nu+n+1} \leq \hat{\eta}_c | \hat{C}_{\nu+1} = y , T>\nu+1\right\} \mathrm{d} \mathcal{P}_{0}^{\mathrm{CU}} (y|x) \nonumber \\
&\overset{(c)}{=} \int_{0}^{\hat{\eta}_c}\mathcal{Q}_{n}^{\mathrm{CU}}(y;q)\mathrm{d}\mathcal{P}_{0}^{\mathrm{CU}} (y|x)
\overset{(d)}{=} \left\{
\begin{aligned}
&\mathcal{Q}_{n}^{\mathrm{CU}}(0;q)\mathcal{P}_{0}^{\mathrm{CU}} (0|x) + \int_{0^+}^{\hat{\eta}_c}\mathcal{Q}_{n}^{\mathrm{CU}}(y;q)\mathrm{d}\mathcal{P}_{0}^{\mathrm{CU}} (y|x),
&& 0\leq x < \omega,\\
&\int_{x + \omega}^{\eta_c}\mathcal{Q}_{n}^{\mathrm{CU}}(y;q)\mathrm{d}\mathcal{P}_{0}^{\mathrm{CU}} (y|x),
&&\omega \leq x < \eta_c,
\end{aligned}
\right. \nonumber \\
&\overset{(e)}{=} \left\{
\begin{aligned}
&\mathcal{Q}_{n}^{\mathrm{CU}}(0;q)\left(1 - \mathrm{e}^{\frac{x - \omega}{1+q}}\right)
+ \frac{1}{1+q}\mathrm{e}^{\frac{x - \omega}{1+q}}
\int_{0}^{\eta_c}\mathcal{Q}_{n}^{\mathrm{CU}}(y;q)\mathrm{e}^{-\frac{y}{1+q}}\mathrm{d}y,
&& 0\leq x < \omega,\\
&\frac{1}{1+q}\mathrm{e}^{\frac{x - \omega}{1+q}}
\int_{x - \omega}^{\eta_c}\mathcal{Q}_{n}^{\mathrm{CU}}(y;q)\mathrm{e}^{-\frac{y}{1+q}}\mathrm{d}y,
&&\omega \leq x < \eta_c,
\end{aligned}
\right.\label{DerivationofCUSUMConditionalQn}
\end{align}
\noindent\rule[0.25\baselineskip]{\textwidth}{1pt}
\vspace{-7mm}
\end{figure*}
where
step $(a)$ follows the law of total probability,
step $(b)$ follows the Markov property of the sequence $\{\hat{C}_t:t\geq 1\}$,
step $(c)$ is obtained by the definition of $\mathcal{Q}_{n}^{\mathrm{CU}}(x;q)$ in \eqref{ConditionalCUSUMQL},
step $(d)$ is because there exists a probability mass at $\hat{C}_{\nu+1} = 0$ if $0\leq x < \omega$, and
step $(e)$ is obtained by using \eqref{ConditionCnp1GivenCn}.
Note that $\mathcal{Q}_{1}^{\mathrm{CU}}(x;q)$ can be obtained by using Lemma \ref{CUSUMSTATISTICCONDITIONCDF}.
Based on $\mathcal{Q}_{1}^{\mathrm{CU}}(x;q)$, $\mathcal{Q}_{2}^{\mathrm{CU}}(x;q)$ can be obtained by inserting $\mathcal{Q}_{1}^{\mathrm{CU}}(x;q)$ into \eqref{DerivationofCUSUMConditionalQn}.
Similarly, for $n > 2$, $\mathcal{Q}_{n}^{\mathrm{CU}}(x;q)$ can be obtained by inserting $\mathcal{Q}_{n-1}^{\mathrm{CU}}(x;q)$ into \eqref{DerivationofCUSUMConditionalQn}. Using mathematical induction we obtain the result.

\subsection{The derivation of Theorem \ref{Theorem:StationaryCUSUMdistribution}}
\label{ProofPRIORCnu}
For $\tilde{\mathcal{G}}_{n}^{\mathrm{CU}}(x;q)$ and $\mathcal{G}_{n-1}^{\mathrm{CU}}(x;q)$ with $n\geq 2$, we have the recursive integral equation \eqref{PriorDistributionCUSUM},
\begin{figure*}
\begin{align}
\tilde{\mathcal{G}}_n^{\mathrm{CU}}(x;q) &\overset{(a)}{=} \int_{0}^{\hat{\eta}_c}
\mathcal{P}_{\infty}^{\mathrm{CU}}(x|y) \mathrm{d} \mathcal{G}_{n-1}^{\mathrm{CU}}(y;q)
\overset{(b)}{=} \mathcal{P}_{\infty}^{\mathrm{CU}}(x|0)\mathcal{G}_{n-1}^{\mathrm{CU}}(0;q) + \int_{0^+}^{\hat{\eta}_c}
\mathcal{P}_{\infty}^{\mathrm{CU}}(x|y) \mathrm{d} \mathcal{G}_{n-1}^{\mathrm{CU}}(y;q) \nonumber \\
&\overset{(c)}{=} \left\{
\begin{aligned}
&\left(1 - \mathrm{e}^{-\left(x+\omega\right)}\right)\mathcal{G}_{n-1}^{\mathrm{CU}}(0;q)  + \int_{0^+}^{\hat{\eta}_c}
\left(1 - \mathrm{e}^{-\left(x+\omega-y\right)}\right)\mathrm{d} \mathcal{G}_{n-1}^{\mathrm{CU}}(y;q),\text{ if }\hat{\eta}_c - \omega \leq x, \\
&\left(1 - \mathrm{e}^{-\left(x+\omega\right)}\right)\mathcal{G}_{n-1}^{\mathrm{CU}}(0;q)+ \int_{0^+}^{x+\omega}
\left(1 - \mathrm{e}^{-\left(x+\omega-y\right)}\right)\mathrm{d} \mathcal{G}_{n-1}^{\mathrm{CU}}(y;q),
\text{ if }0\leq x < \hat{\eta}_c - \omega,
\end{aligned}\right.\nonumber \\
&\overset{(d)}{=} \left\{
\begin{aligned}
&\left(1 - \mathrm{e}^{-\left(x+\omega-\hat{\eta}_c\right)}\right) + \mathrm{e}^{- \left(x+\omega\right)}\int_{0}^{\hat{\eta}_c} \mathcal{G}_{n-1}^{\mathrm{CU}}(y;q)\mathrm{e}^{y} \mathrm{d} y,\quad \text{ if }\hat{\eta}_c - \omega \leq x, \\
&\mathrm{e}^{- \left(x+\omega\right)} \int_{0}^{x+\omega} \mathcal{G}_{n-1}^{\mathrm{CU}}(y;q)
\mathrm{e}^{ y} \mathrm{d}y,\text{ if }0\leq x < \hat{\eta}_c - \omega, \\
\end{aligned}\right.
\label{PriorDistributionCUSUM}
\end{align}
\noindent\rule[0.25\baselineskip]{\textwidth}{1pt}
\vspace{-7mm}
\end{figure*}
where
step $(a)$ follows the law of total probability,
step $(b)$ is because of the probability mass at $\hat{C}_{n-1} = 0$,
step $(c)$ is due to Lemma \ref{CUSUMSTATISTICCONDITIONCDF}, and
step $(d)$ is obtained by using the integral by parts.
We can obtain $\tilde{\mathcal{G}}_{1}^{\mathrm{CU}}(x;q)$ and $\mathcal{G}_{1}^{\mathrm{CU}}(x;q)$ according to Lemma \ref{CUSUMSTATISTICCONDITIONCDF}.
For $n\geq2$, $\tilde{\mathcal{G}}_{n}^{\mathrm{CU}}(x;q)$ can be obtained by inserting $\mathcal{G}_{n-1}^{\mathrm{CU}}(x;q)$ into \eqref{PriorDistributionCUSUM}.

\subsection{Numerically calculation of $\mathcal{Q}_n^{\mathrm{SR}}(x;q)$}
\label{Sec:QnxqNumCal}
Denote $\xi_i$ for $0\leq i\leq N$ satisfying $\frac{1}{1+q}=\xi_0<\xi_1<\xi_2<\cdots<\xi_N=\eta_r$ as $N+1$ sample points in $[0,\eta_r]$. Using trapezoidal quadrature rule, \eqref{RecursiveSRConditional} is approximated as
\begin{align}
\mathcal{Q}_n^{\mathrm{SR}}(\xi_i;q)
\approx&
\frac{1}{2}\sum_{j = 0}^{N-1}
\Big\{ \left( \mathcal{Q}_{n-1}^{\mathrm{SR}}(\xi_j;q) + \mathcal{Q}_{n-1}^{\mathrm{SR}}(\xi_{j+1};q) \right)\nonumber \\
&\quad\quad\quad \times
\left(  \mathcal{P}_{0}^{\mathrm{SR}}\left(\xi_{j+1}|\xi_i\right) -  \mathcal{P}_{0}^{\mathrm{SR}}\left(\xi_j|\xi_i\right) \right) \Big\} \nonumber \\
=& \bm{k}_i^T \bm{t}_{n-1},
\label{QuadratureQnSRxq}
\end{align}
where we denote
\begin{align}
\bm{t}_n \triangleq \left[\mathcal{Q}_n^{\mathrm{SR}}(\xi_0;q),\mathcal{Q}_n^{\mathrm{SR}}(\xi_1;q),\cdots,
\mathcal{Q}_n^{\mathrm{SR}}(\xi_N;q)\right]^T,\nonumber
\end{align}
and $\bm{k}_i$ is a $(N+1)$-dimensional column vector. According to \eqref{QuadratureQnSRxq}, we obtain
\begin{align}
\bm{t}_n \approx  \bm{K}^T \bm{t}_{n-1} \approx
\underbrace{ \bm{K}^T \bm{K}^T \cdots \bm{K}^T }_{n-1\text{ folds}}\bm{t}_{1},\nonumber
\end{align}
where $\bm{K} \triangleq [\bm{k}_0, \bm{k}_1, \cdots, \bm{k}_N]$.

\end{document}